\documentclass[10pt,journal,compsoc]{IEEEtran}
\ifCLASSOPTIONcompsoc
\usepackage[nocompress]{cite}
\else
\usepackage{cite}
\fi

\ifCLASSINFOpdf
\else
\fi
\usepackage{amssymb,amsmath,amsfonts,amsthm}
\usepackage{mathrsfs}
\usepackage{cite}
\usepackage{array}
\usepackage{tabularx}
\usepackage{color}
\usepackage{mathtools}
\usepackage{subfigure}%\usepackage{mathtools}
\usepackage{graphicx}
\usepackage{bbm}
\usepackage{dsfont}
\usepackage{setspace}
\usepackage{multirow}
\DeclareMathOperator{\sinc}{sinc}
\renewcommand{\IEEEQED}{\IEEEQEDopen}

\newtheorem{theorem}{Theorem}
\newtheorem{lemma}{Lemma}

\newtheorem{propi}{Proposition}

\newtheorem{remark}{Remark}
\newcommand{\argmax}{\mathop{\rm arg~max}\limits}

\newcommand{\al}{\frac{\lambda (1-\mu_1)}{(1-\lambda) \mu_1} }
\IEEEoverridecommandlockouts

\begin{document}

\title{On the Performance of Delay Aware Shared Access with Priorities}

\author{Zheng~Chen, Nikolaos~Pappas, Marios~Kountouris and~Vangelis~Angelakis
\IEEEcompsocitemizethanks{\IEEEcompsocthanksitem Z. Chen is with the Laboratoire de Signaux et Syst\`{e}mes (L2S, UMR8506),
		CentraleSup\'{e}lec - CNRS - Universit\'{e} Paris-Sud,
		Gif-sur-Yvette, France. \protect\\
		Email: zheng.chen@centralesupelec.fr. 
		\IEEEcompsocthanksitem N. Pappas and V. Angelakis are with Department of Science and Technology, Link\"{o}ping University, Norrk\"{o}ping, Sweden. \protect\\
		Email: \{nikolaos.pappas, vangelis.angelakis\}@liu.se. 
		\IEEEcompsocthanksitem M. Kountouris is with the Mathematical and Algorithmic Sciences Lab, France Research Center, Huawei Technologies Co. Ltd. \protect\\
		Email: marios.kountouris@huawei.com. % <-this % stops a space
		\IEEEcompsocthanksitem This work was presented in part in \cite{pappas_crowncom} and \cite{wowmom2016}.
		}
}

\IEEEtitleabstractindextext{%
	\begin{abstract}
	In this paper, we analyze a shared access network with a fixed primary node and randomly distributed secondary nodes whose distribution follows a Poisson point process (PPP). The secondaries use a random access protocol allowing them to access the channel with probabilities that depend on the queue size of the primary. Assuming a system with multipacket reception (MPR) receivers having bursty packet arrivals at the primary and saturation at the secondaries, our protocol can be tuned to alleviate congestion at the primary. We study the throughput of the secondary network and the primary average delay, as well as the impact of the secondary node access probability and transmit power. We formulate an optimization problem to maximize the throughput of the secondary network under delay constraints for the primary node, which in the case that no congestion control is performed has a closed form expression providing the optimal access probability. Our numerical results illustrate the impact of network operating parameters on the performance of the proposed priority-based shared access protocol. 
	\end{abstract}

}
\maketitle

\IEEEdisplaynontitleabstractindextext

\IEEEraisesectionheading{\section{Introduction}\label{sec:introduction}}
\IEEEPARstart{M}obile device proliferation is creating tremendous pressure on the capacity of current wireless networks. Due to the scarcity of the radio spectrum, several flexible spectrum management approaches have emerged. Spectrum sharing, licensed-assisted access (LAA), licensed sharing access (LSA), and cognitive radio \cite{zhao:survey} are some novel paradigms providing efficient and flexible spectrum utilization. In a cognitive-inspired shared access network, the unlicensed (secondary) users opportunistically access the under-utilized spectrum of the licensed (primary) network and adjust their transmissions so as not to create harmful interference to the primary user. This network setting can also model underlay device-to-device (D2D) communication in cellular networks, which is seen as a key enabler for 5G mobile communication systems \cite{asadi2014survey, b:fixed-rate, b:machine-type} and the Internet of Things \cite{IoT}.

The conventional access protocol for the secondary node is to vacate the spectrum when the primary node is active, in other words, the secondary node can only be active when the channel is idle in order to avoid collision with the primary transmission. However, due to the imperfect knowledge of the channel occupancy, collisions may be inevitable. Scheduling policies for the secondary user under partial channel state information are developed in \cite{b:jsac, neely}.
Random access protocols with multipacket reception (MPR) are proposed in \cite{b:FanousJSAC2013}, where secondary nodes make transmission attempts with a given probability. Compared to the traditional collision channel model, the MPR channel \cite{ghez:stability, tong:multipacket, b:mpr, b:PappasTWC2015} captures the interference at the physical layer in a more efficient way, because a transmission may succeed even in the presence of interference. 
Nevertheless, spectrum sharing between primary and secondary nodes in MPR channel unavoidably creates interference among concurrent transmissions \cite{cognitive_interference, b:PappasJCN2012, b:KompellaTON2014}. Taking into account the interference caused by the secondary network and affecting the primary user, a judicious access protocol for the secondary node has to be carefully designed so that the quality-of-service (QoS) of the primary user is not degraded.

\subsection{Related Work}
In~\cite{pappas_crowncom}, we analyzed the throughput of the secondary network when MPR capability is enabled in a cognitive network with congestion control on the primary user. Using the collision channel scenario, throughput optimization with deadline constraints on a single secondary user accessing a multi-channel system is studied in \cite{multi-channel}. The optimal stopping rule and power control strategy are provided in terms of closed-form expressions. In \cite{joint} the joint scheduling and power control is considered in order to minimize the sum average secondary delay subject to interference constraints at the primary user. However, prior work has not studied the random access protocol design which takes into account both the throughput of the secondary network and the delay of the primary one.

Most of the prior studies on cognitive radio and shared access networks either assume a single secondary node or multiple secondary nodes in a fixed network topology. To the best of our knowledge, the throughput and delay analysis of a large-scale shared access network with highly mobile secondary uses at random locations has not been reported in the literature. Using tools from stochastic geometry, the secondary node distribution can be modeled as a Poisson point process (PPP) \cite{stogeo_app, b:haenggi2012stochastic}, which is a widely used spatial model for the node distribution in dense wireless networks. Existing results on the interference and outage distribution in PPP networks provide direct connection between the interference level and the node density, thus allowing us to characterize the spatially averaged throughput of the secondary network and interference as a function of the secondary node density. Therefore, the primary average delay can be well confined by adjusting the access probability of the secondary nodes in the random access protocol.

\subsection{Contribution}
This work extends and enhances our early works in \cite{pappas_crowncom, wowmom2016} in the following aspects. 
\begin{itemize}
	\item We propose a delay-aware shared access network with congestion control in the primary network. A large-scale secondary network is considered in which the nodes are distributed according to a stochastic point process. 	
	\item We derive the average queue size and delay of the primary user as function of the secondary node access probability and transmit power.
	\item We introduce an optimization problem to maximize the throughput of the secondary network subject to the delay constraints on the primary user. We analyze the impact of different network parameters on the throughput and delay behavior of our studied network. 
	\item For the particular case with no congestion control, we provide closed-form expressions for the optimal access probability of the secondary nodes. The analytical results are shown through simulations to be very accurate, allowing us to optimize the performance of a large-scale shared access network with simple control schemes. For the case with congestion control, we evaluate with numerical methods the optimal solution for our shared access protocol design.
\end{itemize}

\section{System Model} 
\label{sec:model}
\subsection{Network Topology}
\begin{figure}[h]
	\includegraphics[scale=0.25]{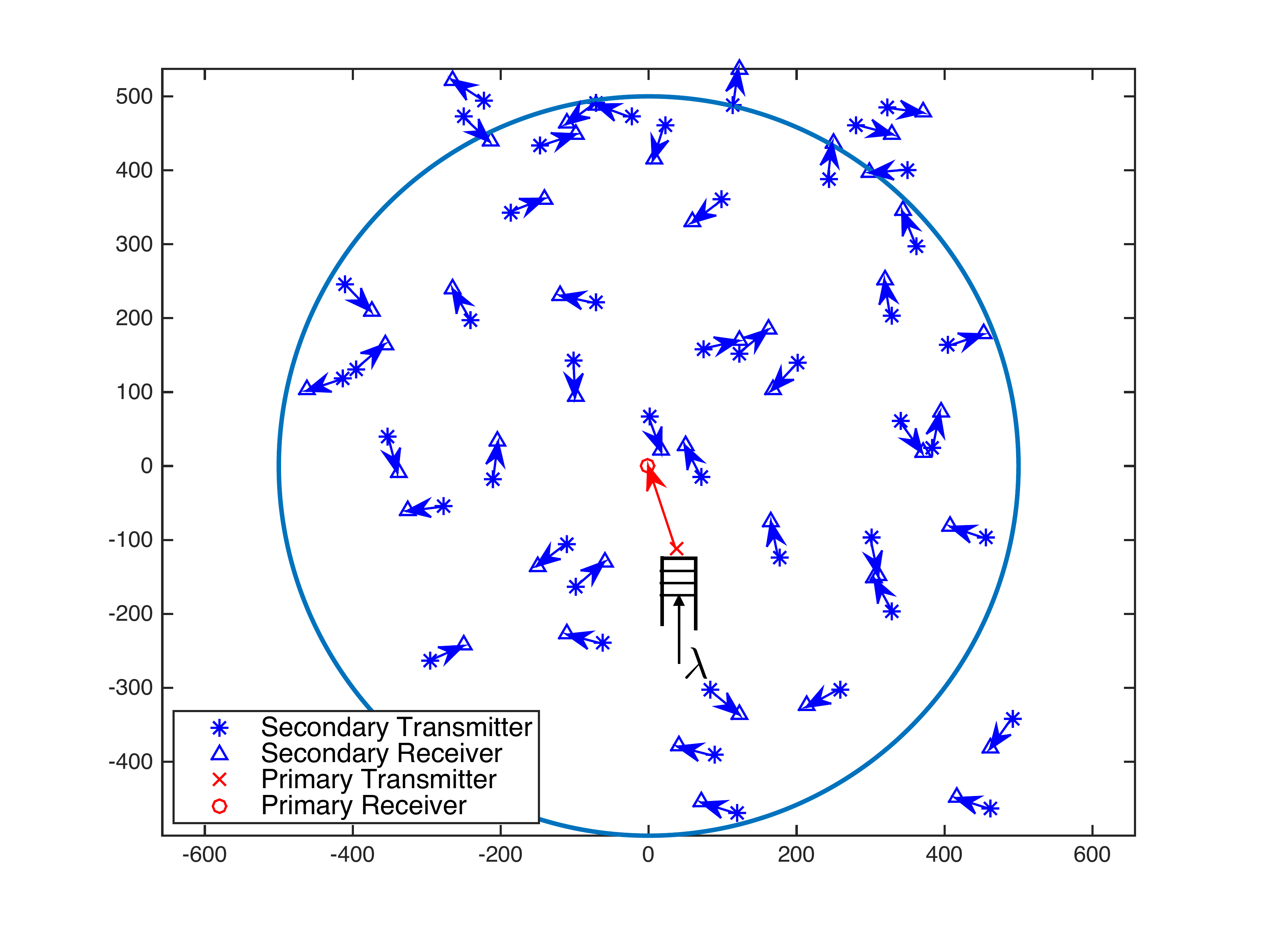}
	\caption{Shared access network topology: one primary receiver centered at the origin with PPP distributed secondary transmitters under a given density (here $\lambda_s=5\times10^{-5}$). }
	\label{fig:system_model}
\end{figure}

We consider a shared access network, in which one primary source-destination pair and many secondary communication pairs share the same spectrum, as shown in Fig.\ref{fig:system_model}. The network region we study is a circular disk $\mathcal{C}$ with radius $R$. The primary receiver is centered at the origin of $\mathcal{C}$. The primary transmitter is located at fixed location with distance $d_p$ to the primary receiver, which is common in infrastructure-based communication.
We assume that the secondary transmitters are distributed in the two-dimensional Euclidean plane $\mathbb{R}^2$ according to a homogeneous Poisson point process (PPP) $\Phi_s=\{x_i \in \mathbb{R}^2, \forall i\in \mathbb{N}^{+}\}$ with intensity $\lambda_s$, where $x_i$ denotes the location of the $i$-th secondary transmitter. Their associated receivers are distributed at isotropic directions with fixed distance $d_s$ from their transmitters. For each realization of the PPP, the number of secondary transmitters in our network region $\mathcal{C}$ is a Poisson random variable with mean value $\lambda_s \pi R^2$. The time is slotted and each packet transmission occupies one time slot. We assume that the receivers have multipacket reception (MPR) capabilities and that the secondary nodes can transmit simultaneously with the primary node \cite{MPR}. 

The primary source has an infinite capacity queue $Q$ for storing arriving packets of fixed length. The arrival process at the primary transmitter is modeled as a Bernoulli process with average rate $\lambda$ packets per slot. The secondary node queue is assumed to be saturated, i.e., it always has a packet waiting to be transmitted.

\subsection{Priority Based Protocol Model}
\label{sec:protocol_model}
We consider the following priority-based protocol, which is an extension of that proposed in~\cite{pappas_crowncom}.  The primary node transmits a packet whenever backlogged, while the secondary nodes access the channel with a probability that depends on the queue size of the primary node, such that will not deteriorate the performance of the primary user.
Denote $Q$ the queue size in the primary node, the activity of the primary and secondary transmitters in a time slot are controlled in the following cases:
\begin{itemize}
	\item \textit{Case 1}: When $Q=0$, the primary transmitter does not have packet to transmit, thus remains silent. Secondary transmitters randomly access the channel with probability $q_1$.
	\item \textit{Case 2}:  When $1 \leq Q\leq M$, the primary transmitter transmits one packet. Secondary transmitters randomly access the channel with probability $q_2$.
	\item  \textit{Case 3}: When $Q> M$, the primary transmitter transmits one packet. Secondary transmitters remain silent.
\end{itemize}
For brevity we use PT and PR to denote the primary transmitter and receiver respectively, and ST and SR for denoting the secondary transmitter and receiver.

The threshold $M$ plays the role of a congestion limit for the primary node, meaning that when the queue reaches this size, then the STs do not attempt to transmit any packet. When $M=\infty$, the protocol model is simplified to the case without congestion control. 

Note that we use two random access probabilities for the secondary nodes because the SRs experience different interference levels depending on whether the PT is active or not. Thus, the optimal access probabilities in these two cases need to be investigated separately.

\section{Physical Model Successful Transmission Analysis}
\label{sec:phy_model}
The MPR physical model is a generalized form of the packet erasure model. At the receiver side, a packet can be decoded correctly by the receiver if the received signal-to-interference-plus-noise ratio (SINR) exceeds a prescribed threshold $\theta$. Given a set $\mathcal{T}$ of nodes transmitting during the same time slot, the received SINR at the $i$-th receiving node is given by
\begin{equation*}
{\rm SINR}_i=\frac{P_{i}|h_{i,i}|^2 d_{i,i}^{-\alpha}}{\sum_{j\in \mathcal{T}\backslash\left\{i\right\}} P_{j}|h_{j,i}|^2 d_{j,i}^{-\alpha}+\sigma^2},
\end{equation*}
where $P_{i}$ denotes the power of the transmitting node $i$; $h_{j,i}$ denotes the small-scale channel fading from the transmitter $j$ to the receiver $i$, which follows $\mathcal{CN}(0,1)$ (Rayleigh fading); $d_{j,i}$ denotes the distance between the transmitter $j$ to the receiver $i$. Here we assume a standard distance-dependent power law pathloss attenuation $d^{-\alpha}$, where $\alpha>2$ denotes the pathloss exponent. $\sigma^2$ denotes the background noise power. 

Let $P_1$ and $P_2$ be the transmit powers of the PT and the STs, respectively. 
In the following we refer to the primary node by node $0$, while the secondary nodes are labeled with index $i\geq 1$. 
Denote $x_0$ the location of the PT and recall that the distribution of the STs is given by $\Phi_s$, then we have $\mathcal{T}\subseteq\{x_0\cup \Phi_s\}$. Note that in this work when we refer to the set of locations of the transmitting nodes, it means the set of transmitting nodes at these locations.
 
Following the description of our access protocol presented in Section~\ref{sec:protocol_model}, to derive the success probability of the primary and secondary nodes we need to consider three cases. 

\subsection{Case 1}
When $Q=0$, the PT is silent and the STs attempt packet transmission with probability $q_1$. Denote $\Phi_a^1$ the locations of active STs, as a result of independent thinning \cite{b:haenggi2012stochastic}, $\Phi_a^1$ follows a homogeneous PPP with intensity $q_1 \lambda_s$. Hence, we have the active transmitter set as $\mathcal{T}=\Phi_a^1$.
 
Without loss of generality, we consider an arbitrary (typical) active secondary pair $i$ in our network region. Denote $p_{2/2}$ the success probability of the typical secondary pair when only the STs from $\Phi_a^1$ are active, we have 
	\begin{align}
	p_{2/2}&=\mathbb{P}[\rm SINR_i >\theta \mid \mathcal{T}=\Phi_a^1]  \nonumber\\
	&=\mathbb{P}\left[\frac{P_2|h_{i,i}|^2 d_s^{-\alpha}}{\sigma^2+\sum_{j\in \Phi_a^1\backslash\left\{i\right\}} P_{2}|h_{j,i}|^2 d_{j,i}^{-\alpha}} >\theta\right]  \nonumber\\
	&\mathop{=}\limits^{(a)} \exp\left(-\frac{\pi q_1\lambda_s d_s^2 \theta^{\frac{2}{\alpha}}}{\sinc(2/\alpha)}\right) \exp\left(-\frac{\theta \sigma^2 d_s^\alpha}{P_2}\right).
	\label{eq:p22}
	\end{align}
Here, $(a)$ comes from $|h_{i,i}|^2 \sim \exp(1)$ and the probability generating functional (PGFL) of the PPP \cite{haenggi_interference}. For a specific realization of the PPP, $p_{2/2}$ represents the percentage of active secondary pairs having successful transmission. It can also be seen as the probability of the typical active secondary pair to have successful transmission, averaging over different realizations of the PPP.

\subsection{Case 2} When  $1\leq Q\leq M$, both the PT and part of the STs are active. Similarly, with independent thinning probability $q_2$, the locations of active STs follow another homogeneous PPP, denoted by $\Phi_a^2$, with intensity $q_2\lambda_s$. In that case, the active transmitter set contains both the PT and the active STs, i.e., $\mathcal{T}=\{x_0 \cup \Phi_a^2\}$.

Denote $p_{1/1, 2}$ and $p_{2/1, 2}$ the success probabilities of the primary and secondary pairs when both types of nodes are active. With the help of existing results on the interference and outage distribution in PPP networks \cite{b:haenggi2012stochastic}, we have the success probability of the primary transmission when the secondary network is active, given as
\begin{align}
p_{1/1, 2}&=\mathbb{P}\left[\rm SINR_0 >\theta \mid \mathcal{T}=\{x_0 \cup \Phi_a^2\}\right]  \nonumber\\
&=\mathbb{P}\left[\frac{P_1|h_{0,0}|^2 d_{p}^{-\alpha}}{\sigma^2+\sum_{j\in \Phi_a^2} P_{2}|h_{j,0}|^2 d_{j,0}^{-\alpha}} >\theta\right]  \nonumber\\
&=\exp\left[-\frac{\pi q_2 \lambda_s \left(\theta\frac{P_2}{P_1}\right)^{2/\alpha}d_p^2}{\sinc(2/\alpha)}\right] \exp\left(-\frac{\theta \sigma^2 d_p^\alpha}{P_1}\right). 
\label{eq:p112}
\end{align}

For the active secondary nodes, considering an arbitrary (typical) active secondary pair $i$,
we obtain the success probability in the following proposition.
 \begin{propi}
 	\label{propi_p212}
 	The success probability of the typical secondary pair, when the active transmitters are $\mathcal{T}=\{x_0 \cup \Phi_a^2\}$, is given by
 	\begin{align}
 	p_{2/1, 2}
 	\simeq\exp\left[-\frac{\pi q_2\lambda_s d_s^2 \theta^{\frac{2}{\alpha}}}{\sinc(2/\alpha)}\right]\frac{\exp\left(-\frac{\theta \sigma^2 d_s^\alpha}{P_2}\right)}{1+\frac{d_s^2}{\mathbb{E}[d_{0,i}]^2}\left(\theta\frac{P_1}{P_2}\right)^{\frac{2}{\alpha}}},
 	\label{eq:p212}
 	\end{align}
 	where $\mathbb{E}[d_{0,i}]=\int_{0}^{2\pi}\frac{1}{2\pi}\int_{0}^{R}\frac{2 r}{R^2}\sqrt{r^2+d_p^2-2r d_p \cos \varphi}\text{d}r\text{d}\varphi$. 
 \end{propi}
 
 \begin{IEEEproof}
 	\textnormal{See Appendix \ref{appen1}.}
 \end{IEEEproof}

\subsection{Case 3} When $Q>M$, only the PT is active. Denote $p_{1/1}$ the success probability of the primary pair when all the STs are silent, we have 
	\begin{align}
	p_{1/1}&=\mathbb{P}[\rm SINR_0 >\theta \mid \mathcal{T}=x_0 ]  =\mathbb{P}\left[\frac{P_1|h_{0,0}|^2 d_{p}^{-\alpha}}{\sigma^2} >\theta\right]  \nonumber\\
	&= \exp\left(-\frac{\theta \sigma^2 d_p^\alpha}{P_1}\right). 
	\label{eq:p11}
	\end{align}

Note that $p_{1/1} > p_{1/1,2}$ and $p_{2/2} > p_{2/1,2}$ always hold.

\section{Network Performance Metrics}
\label{sec:def_metric}
In this section, we define several relevant metrics for the performance evaluation of the proposed priority-based protocol with congestion control.
\subsection{Throughput of the Secondary Network}
\label{sec:def_throughput}
For the considered shared access network, we aim at evaluating the throughput of the secondary network, \textit{abbreviated as secondary throughput}, which is the number of packets per slot that can be successfully transmitted by the active secondary nodes to their destinations.
In order to be consistent with the PPP model where the secondary nodes are generated with a certain density $\lambda_s$, we define the secondary throughput as the throughput of the secondary network per unit area, given as
\begin{equation}
T_s = \lambda_s \mathbb{P}[\rm SINR_{i\in\Phi_s}>\theta]. 
\end{equation}
Recall that the active STs is with density $q_1\lambda_s$ when the primary queue is empty, i.e., $Q=0$. When the primary queue is $1\leq Q \leq M$, then the active STs have density $q_2 \lambda_s$. Hence, we have 
\begin{align}
T_s = &\mathbb{P}[Q=0] \cdot q_1\lambda_s  \mathbb{P}[\rm SINR_{i\in\Phi_a^1 } >\theta \mid Q=0]  \nonumber \\
&+ \mathbb{P}[1 \leq Q\leq M]  \cdot q_2 \lambda_s \mathbb{P}[\rm SINR_{i\in\Phi_a^2} >\theta \mid 1 \leq Q\leq M]  \nonumber \\
=& \lambda_s\left\{\mathbb{P}[Q=0] \cdot q_1 p_{2/2}+\mathbb{P}[1 \leq Q\leq M] \cdot q_2 p_{2/1,2}\right\},
\label{eq:def_sd_throughput}
\end{align} 
where $p_{2/2}$ and $p_{2/1,2}$ are given in \eqref{eq:p22} and \eqref{eq:p212}, respectively.

\subsection{Primary Service Rate}
The service rate of the primary given a certain SINR target can be defined as the percentage of successfully transmitted packets per time slot. Dividing the cases by the primary queue size greater or less than $M$, when $1\leq Q \leq M$, we have the primary service rate given by  
\begin{equation}
\mu_1=p_{1/1,2}.
\end{equation}
When $Q>M$, the service rate is
\begin{equation}
\mu_2=p_{1/1}.
\end{equation}
Combining the two cases, we have the average service rate of the primary, denoted by $\bar{\mu}$, given by
\begin{equation}
\bar{\mu}
=\frac{\mathbb{P}[1\leq Q \leq M] \mu_1 +\mathbb{P}[Q > M] \mu_2}{\mathbb{P}[Q\geq 1]}.
\end{equation}

\subsection{Primary Average Delay}
\label{sec:def_delay}
The delay per packet at the primary node consists of the queueing delay and the transmission delay from the PT to the PR. From Little's law, we obtain the queueing delay which is related to the average queue size per packet arrival. The transmission delay is inversely proportional to the average service rate \cite{b:Bertsekas}.

Denote $\bar{D}_p$ the primary average delay per packet, we have
\begin{equation}
\bar{D}_p =\frac{\bar{Q}}{\lambda} +\frac{1}{\bar{\mu}} ,   
\label{eq:delay_expression}     
\end{equation}
where $\bar{Q}$ and $\bar{\mu}$ are the average queue size and the average service rate of the primary, which will be analyzed with closed-form expressions in Section~\ref{sec:Analysis}.

\section{Analysis of the Primary Queue and Delay} 
\label{sec:Analysis}
From the definition of the metrics in Section~\ref{sec:def_metric}, we see that the secondary throughput and the primary delay depends on the state of the primary queue size. Therefore, we need to derive first $\mathbb{P}\left[Q=0\right]$ and $\mathbb{P}\left[1 \leq Q \leq M \right]$. 

\begin{figure}[h]
\centering
\includegraphics[scale=0.35]{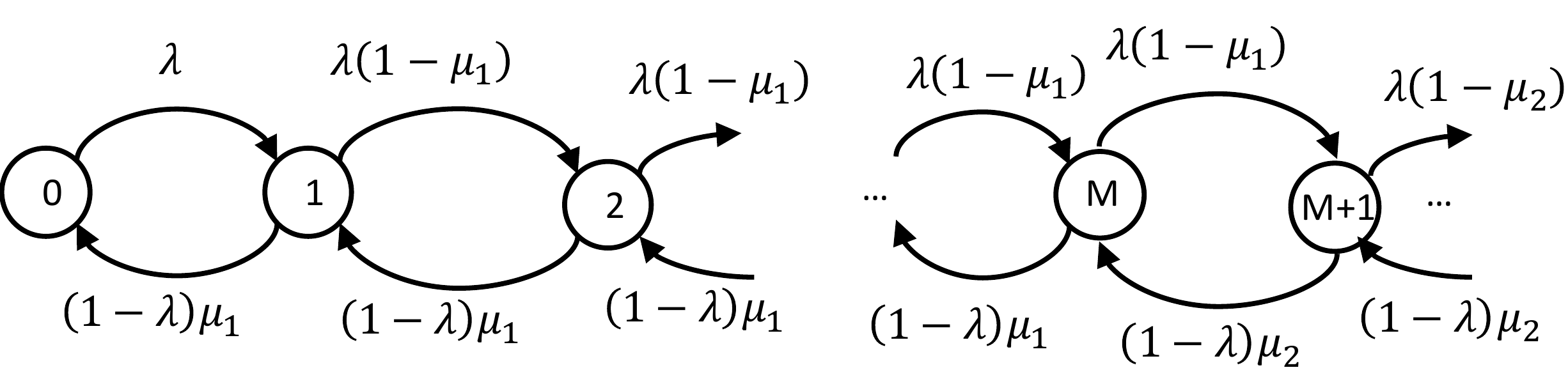}
\centering
\caption{The Discrete Time Markov Chain which models the queue evolution at the primary node.}
\label{fig:dtmc}
\end{figure}

We model the primary queue as a discrete time Markov Chain (DTMC), which describes the queue evolution and is presented in Fig.~\ref{fig:dtmc}. Each state is denoted by an integer and represents the queue size. The packet arrival rate is always $\lambda$. The service rate is $\mu_1=p_{1/1, 2}$ when $1 \leq Q \leq M$, and is $\mu_2=p_{1/1}$ when $Q>M$. From our analysis in Section~\ref{sec:phy_model}, we know that $\mu_2>\mu_1$. All the metrics related to the rate are measured by the average number of packets per time slot. 

Denote $\pi$ the stationary distribution of the DTMC, where $\pi(i) = \mathbb{P}\left[Q=i \right]$ is the probability that the queue has $i$ packets in its steady state. We have the following lemma.

\begin{lemma}
	\label{lemma_dtmc}
	The stationary distribution of the DTMC described in Fig.~\ref{fig:dtmc} is given in the following cases:
	\begin{itemize}
		\item For $1\leq Q\leq M$, we have
		\begin{equation}
		\pi(i) = \frac{\lambda^i (1-\mu_1)^{i-1}}{(1-\lambda)^i \mu_1^i} \pi(0);
		\label{eq:pi}
		\end{equation}
		\item For $i>M$, we have
		\begin{equation}
		\pi(i) = \frac{\lambda^{i} (1-\mu_1)^M (1-\mu_2)^{i-M-1}}{(1-\lambda)^{i} \mu_1^{M} \mu_2^{i-M}} \pi(0), 
		\label{eq:pi_bis}
		\end{equation}
	\end{itemize}
		where $\pi(0)$ is the probability that the queue is empty, given by
		\begin{equation}
		\pi(0)= \left\{
		\begin{array}{rcl}
		& \frac{(\mu_1 - \lambda)(\mu_2 - \lambda)}{\mu_1 \mu_2 - \lambda \mu_1 - \lambda \left[\al\right]^M (\mu_2 - \mu_1) }  & \text{if}~~\lambda\neq\mu_1\\
  &&\\
		&\frac{\mu_2-\mu_1}{\mu_1+(\mu_2-\mu_1)\frac{M+1-\mu_1}{1-\mu_1}}   &\text{if}~~\lambda=\mu_1.
		\end{array} \right.
		\label{eq:Pempty}
		\end{equation}		
		The queue is stable if and only if $\lambda<\mu_2$.
\end{lemma}
 \begin{IEEEproof}
 	\textnormal{See Appendix \ref{appen2}.}
 \end{IEEEproof}

\begin{remark}
	If without congestion control, the service rate is always $\mu_1$. Obviously the condition to have stable queue is $\lambda<\mu_1$. The congestion control threshold $M$ increases the queue stability region to $\lambda<\mu_2$, implying that the maximum allowed arrival rate at the PT becomes higher. On the other hand, less opportunity will be given to the secondary nodes to be active, because no secondary transmission is allowed when the primary queue size exceeds $M$.   
\end{remark}

In order to simplify the equations, we define $\xi\triangleq\al$. In the remainder of this work we will assume that $\lambda\neq\mu_1$, however the general expressions of our results hold also for $\lambda=\mu_1$, but one should replace the $\pi(0)$ with the corresponding expression in this case.

Based on the results in Lemma~\ref{lemma_dtmc}, we have the following probabilities related to the primary queue size.

\begin{lemma}
	\label{lemma_p1m}
	When the primary queue is stable and $\lambda\neq\mu_1$, the probability to have $1 \leq Q \leq M$ is 
	\begin{equation}
	\mathbb{P}\left[1 \leq Q \leq M\right]= \frac{\lambda \left(1-\xi^M \right)(\mu_2 - \lambda)}{\mu_1 \mu_2 - \lambda \mu_1 - \lambda \xi^M (\mu_2 - \mu_1)}.
	\label{eq:Pr_1M}
	\end{equation}
	The probability to have $Q>M$ is
	\begin{equation}
	\mathbb{P}\left[Q > M\right]= \frac{\lambda \xi^M  (\mu_1-\lambda)}{\mu_1 \mu_2 - \lambda \mu_1 - \lambda \xi^M (\mu_2 - \mu_1)}.
	\label{eq:Pr_M}
	\end{equation} 
\end{lemma}
 \begin{IEEEproof}
 	\textnormal{See Appendix \ref{appen3}.}
 \end{IEEEproof}

We give the average queue size and the average delay of the primary in the following theorem. 
\begin{theorem}
	\label{lemma_delay}
	The average queue size of the primary is given by
	\begin{equation}
	\bar{Q}=\frac{N_1+N_2}{\mu_1\mu_2 -\lambda\mu_1 -\lambda \xi^M  (\mu_2-\mu_1) },
	\end{equation}
	where 
	\begin{equation}
	N_1=\lambda(1-\lambda)\mu_1 \frac{\mu_2-\lambda}{\mu_1-\lambda}\left[M \xi^{M+1}-\xi^M(M+1) +1\right],
	\end{equation}
	and 
	\begin{equation}
	N_2=\xi^M \lambda(\mu_1-\lambda)\left[M+\frac{(1-\lambda)\mu_2}{\mu_2-\lambda}\right].
	\end{equation}
		
	The primary average delay is given by
	\begin{equation}
	\bar{D}_p=\frac{\bar{Q}}{\lambda} +\frac{\mu_2-\lambda-\xi^M(\mu_2-\mu_1)}{\left(1-\xi^M \right)(\mu_2 - \lambda) \mu_1+\xi^M  (\mu_1-\lambda) \mu_2}.   
	\label{eq:delay_epression}  
	\end{equation}
\end{theorem}
\begin{IEEEproof}
	\textnormal{See Appendix \ref{appen4}.}
\end{IEEEproof}

\begin{remark}
For a certain packet arrival rate $\lambda$ at the PT, $\bar{D}_p$ is independent of $q_1$. The primary queue size augments with $q_2$ because of the lower service rate $\mu_1$, which leads to higher queueing delay. The transmission delay also increases with $q_2$. As a result, $\bar{D}_p$ is an increasing function of $q_2$. Similarly, we know that $\bar{D}_p$ also increases with $P_2$.
\end{remark}

\section{Secondary Throughput Optimization with Primary Delay Constraints}
In our considered shared access network, spectrum sharing between the primary and secondary users can be exploited in order to bring secondary throughput gains at the expense of increasing interference to the PR. In order to protect the QoS of the primary user, the secondary interference must be kept below a certain level, which corresponds to the thresholds on the ST access probability $q_2$ and transmit power $P_2$. 

In this section, we analyze the secondary throughput as a function of $q_2$ and $P_2$ with respect to the primary delay constraints. 

\subsection{General Case}
From the definition of the secondary throughput in \eqref{eq:def_sd_throughput}, with the help of the results in Lemma~\ref{lemma_dtmc} and Lemma~\ref{lemma_p1m}, we have
\begin{equation}
\begin{split}
T_s&=\lambda_s\left(\mathbb{P}[Q=0] \cdot q_1 p_{2/2}+\mathbb{P}[1 \leq Q\leq M] \cdot q_2 p_{2/1,2}\right) \\
&=\lambda_s\frac{(\mu_2 - \lambda) \left[q_1 p_{2/2} (\mu_1 - \lambda)+q_2 p_{2/1,2}\lambda \left(1-\xi^M \right)\right]}{\mu_1 \mu_2 - \lambda \mu_1 - \lambda \xi^M (\mu_2 - \mu_1)}  \\
&=\lambda_s\frac{(p_{1/1} \!-\! \lambda) \left[q_1 p_{2/2} (p_{1/1,2} \!-\!\lambda)\!+\! q_2 p_{2/1,2}\lambda \left(1\!-\!\xi^M \right)\right]}{p_{1/1,2} p_{1/1} - \lambda p_{1/1,2} - \lambda \xi^M (p_{1/1} - p_{1/1,2})}. 
\end{split}
\label{eq:sd_thpt_proba}
\end{equation}

Considering the secondary throughput $T_s$ as a function of the access probability $q_1$, it is obvious that there exists an optimal value $q_1^*=\argmax_{q_1\in[0,1]}~T_s$, which is equivalent to $q_1^*=\argmax_{q_1\in[0,1]}~q_1 p_{2/2}$, where $p_{2/2}$ is given in \eqref{eq:p22}. 
From \cite{b:bacelli, jeff_d2d} we have that the optimal access probability $q_1$ of the STs when the PT is silent is given by
\begin{equation}
q_1^*=\min\left\{\frac{\sinc (\frac{2}{\alpha})}{\pi \lambda_s \theta^{\frac{2}{\alpha}}d_s^2}, 1 \right\},
\label{eq:optimal_q1}
\end{equation} 
which depends only on the ST density $\lambda_s$, secondary link distance $d_s$ and the pathloss exponent $\alpha$. Setting $q_1^*$ in \eqref{eq:sd_thpt_proba}, when the PT transmit power $P_1$ and the packet arrival rate $\lambda$ are fixed, the secondary throughput depends only on the access probability $q_2$ and the transmit power $P_2$.

As mentioned in Section~\ref{sec:Analysis}, the primary average delay is an increasing function of $q_2$ and $P_2$. When $\lambda<\mu_2$, i.e., the primary queue is stable, the delay constraints of the primary user can be translated to the feasible region of the two variables $(q_2, P_2)$, defined as 
\begin{equation}
\mathcal{R}_\mathcal{F}=\{(q_2, P_2): \bar{D}_p<D_{\max}\}.
\label{eq:def_RF_general}
\end{equation}
where $D_{\max}$ is the threshold of the primary average delay . 

In order to achieve the maximum secondary throughput while keeping the primary average delay below the threshold, we formulate the following optimization problem:
\begin{equation}
\centering
(q_2^*, P_2^*)=\argmax_{(q_2, P_2) } ~T_s,
\end{equation}
subject to
\begin{eqnarray}
 \label{condition 1} q_2& \in& [0,1],  \nonumber\\
 \label{condition 2}  P_2 &\in& (0, P_{2, \max}], \nonumber\\
 \label{condition 2} \bar{D}_p(q_2, P_2)&<&D_{\max}, \nonumber
\end{eqnarray} 
where $P_{2, \max}$ is the maximum available power for a ST.

Due to the complexity of the analytical results related to the primary queue, it is difficult to solve the above optimization problem in closed form. Hence, first we investigate the particular case without congestion control, i.e., $M=\infty$. The solution to the optimization problem in the general case is evaluated numerically in Section~\ref{sec:Results}.

\subsection{Case with no Congestion Control ($M=\infty$)}
Without congestion control, the activity of the primary and secondary nodes is simplified into two cases:
\begin{itemize}
	\item When $Q=0$, the PT remains silent. STs randomly access the channel with probability $q_1$. 
	\item  When $Q\geq 1$, the PT transmits one packet. STs randomly access the channel with probability $q_2$.
\end{itemize}

Following the primary queue analysis in Lemma~\ref{fig:dtmc}, we have the probability to have $i$ packets in the primary queue when it is in the steady state, given as
\begin{equation}
\pi(i) = \frac{\lambda^i (1-\mu_1)^{i-1}}{(1-\lambda)^i \mu_1^i} \pi(0),
\end{equation}
where
\begin{equation}
\pi(0)=\mathbb{P}[Q=0]=1-\frac{\lambda}{\mu_1}=1-\frac{\lambda}{p_{1/1,2}}.
\end{equation}
The primary queue is stable if and only if $\lambda<\mu_1$. Thus the feasible region of $(q_2, P_2)$ is defined by
\begin{equation}
\mathcal{R}_\mathcal{F}=\{(q_2, P_2): \bar{D}_p<D_{\max}, p_{1/1,2} >\lambda\},
\end{equation}
The secondary throughput becomes
\begin{align}
T_{s}&=\lambda_s\left(\mathbb{P}[Q=0] \cdot q_1 p_{2/2}+\mathbb{P}[Q\geq 1] \cdot q_2 p_{2/1,2}\right) \nonumber\\
&=\lambda_s\left[\left(1-\frac{\lambda}{p_{1/1,2}}\!\right)\!q_1 p_{2/2}\!+\!\frac{\lambda}{p_{1/1,2}} q_2 p_{2/1,2}\right]\!.
\label{eq:thp_SU_noM}
\end{align}
It is straightforward that the optimal value of $q_1$ is the same as in the case with congestion control, given in \eqref{eq:optimal_q1}. 
Inserting $q_1^{*}$ in \eqref{eq:thp_SU_noM} and denoting $c^{*}=q_1^{*} p_{2/2}(q_1^{*})$ the optimal per-node secondary throughput when $Q=0$, the secondary throughput $T_s$ can be written as a function of $q_2$ as follows
\begin{align}
T_{s}& =\lambda_s\left[c^{*} \left(1-\frac{\lambda}{p_{1/1,2}(q_2)}\right) + \frac{\lambda}{p_{1/1,2}(q_2)} q_2 p_{2/1,2}(q_2)\right]  \nonumber \\
&=\lambda_s\left[c^{*}+\frac{\lambda}{p_{1/1,2}(q_2)}\left(q_2 p_{2/1,2}(q_2)-c^{*} \right)\right].  
\label{eq:thp_SU_noM_2}
\end{align}

Our objective is to find the optimal access probability $q_2^*$ that maximizes the secondary throughput for fixed $P_2$ under the primary delay constraints. For that, the optimization problem is redefined as follows.
\begin{equation}
\centering
q_2^*=\argmax_{q_2 } ~T_s,
\label{eq:optimization}
\end{equation}
subject to
\begin{eqnarray}
 \label{condition 1} q_2& \in& [0,1],   \nonumber \\
\label{condition 2} \bar{D}_p(q_2)&<&D_{\max} ,\nonumber  \\
\label{condition 3} p_{1/1,2} (q_2)&>&\lambda.\nonumber 
\end{eqnarray}

The following lemma provides the global optimal value of $q_2$ without considering the primary delay constraints.
\begin{lemma}
	\label{theo_optimal_qo}
		When $\frac{P_2}{P_1}<\left(d_s/d_p\right)^{\alpha}$ is verified, the global optimal value of $q_2\in [0,1]$ that maximizes the secondary throughput in \eqref{eq:thp_SU_noM_2} is given by
			\begin{equation}
			q_2^{\text{o}}=\min\!\left\{\!\left[-\frac{W\!\left(\!\frac{\lambda_s \kappa_{1}\kappa_{2} c_{12}^{*}}{\kappa_{1}-\kappa_{2}} e^{\frac{\kappa_1}{\kappa_{1}-\kappa_{2}}}\!\right)}{\lambda_s \kappa_1}+\frac{1}{\lambda_s(\kappa_{1}\!-\!\kappa_{2})}\right]^{+}\!\!\!, 1\right\}\!,
			\label{eq:optimal_q2_general}
			\end{equation}
			where $W$ denotes the Lambert W function, $[z]^{+}=\max\{z, 0\}$. $\kappa_1=\frac{\pi d_s^2 \theta^{2/\alpha}}{\sinc(2/\alpha)}$, $\kappa_2=\frac{\pi d_p^2 \left(\theta\frac{P_2}{P_1}\right)^{2/\alpha}}{\sinc(2/\alpha)}$, and $c_{12}^{*}=q_1^{*} p_{2/2}(q_1^{*})\left[1+\frac{d_s^2}{\mathbb{E}[d_{0,i}]^2}\left(\theta\frac{P_1}{P_2}\right)^{\frac{2}{\alpha}}\right]$ are constant parameters related to the network setting.
\end{lemma}
\begin{IEEEproof}
	\textnormal{See Appendix \ref{appen5}.}
\end{IEEEproof}

\begin{remark}
	The value of ST power $P_2$ has a significant impact to the global optimal value of $q_2$. When $P_2$ is very high, the primary transmission can be severely harmed by excess interference. We assume here the practically relevant constraint that $P_2$ satisfies $\frac{P_2}{P_1}<\left(d_s/d_p\right)^{\alpha}$. This choice not only simplifies our analysis on the throughput optimization, but also reflects the evolution of wireless networks in deployments where D2D/M2M communication with very low power nodes could coexist with the traditional high-rate mobile users \cite{andrews2014will}.
\end{remark}

The average queue size of the primary in this case is
\begin{equation}
\bar{Q}=\sum\limits_{i=0}^{+\infty} i \pi(i) =\frac{\lambda(1-\lambda)}{\mu_1 -\lambda}.
\end{equation}
The primary average delay is thus given by 
\begin{equation}
\bar{D}_p =\frac{\bar{Q}}{\lambda}+\frac{1}{\mu_1} = \frac{1-\lambda}{\mu_1-\lambda} +\frac{1}{\mu_1}.
\end{equation}
Then we obtain the feasible region of $(q_2, P_2)$ in the following lemma. 
\begin{lemma}
	\label{theo_feasible_region}
	With respect to the maximum average delay $D_{\max}$ of the primary and the queue stability condition, the feasible region of $(q_2, P_2)$ is given by 
	\begin{equation}
	\mathcal{R}_{\mathcal{F}}\triangleq\left\{(q_2, P_2): q_2< \min\left\{ \frac{\ln (p_{1/1}/\lambda)}{ \lambda_s \kappa_2},  \frac{\ln (p_{1/1}/\eta_1)}{ \lambda_s \kappa_2}\right\}\right\},
	\label{eq:feasible_region_theo}
	\end{equation}
	where $\eta_1=\frac{(D_{\max}-1)\lambda+2+\sqrt{(D_{\max}-1)^2\lambda^2-4\lambda +4}}{2D_{\max}}$, $\kappa_2$ is defined in Lemma~\ref{theo_optimal_qo}.
\end{lemma}
\begin{IEEEproof}
	\textnormal{See Appendix \ref{appen6}.}
\end{IEEEproof}

Theorem~\ref{theo_optimal_q2} provides the optimal $q_2$ which maximizes $T_s$ within the feasible region $\mathcal{R}_{\mathcal{F}}$, as the solution to the optimization problem defined in \eqref{eq:optimization}.

\begin{theorem}
	\label{theo_optimal_q2}
	The optimal access probability $q_2^{*}$ that maximizes the secondary throughput under primary delay constraints is given by
	\begin{equation}
	q_2^{*}=\min\left\{q_2^{\text{o}}, \frac{\ln (p_{1/1}/\lambda)}{ \lambda_s \kappa_2}, \frac{\ln (p_{1/1}/\eta_1)}{ \lambda_s \kappa_2} \right\},
	\end{equation}
	where $q_2^{\text{o}}$ is given in \eqref{eq:optimal_q2_general}, $\eta_1$ is defined in Lemma~\ref{theo_feasible_region}. 
\end{theorem}
\begin{IEEEproof}
	\textnormal{See Appendix \ref{appen7}.}
\end{IEEEproof}

\section{Numerical Results}\label{sec:Results}
\begin{table}[t]
	\centering
	\caption{System Parameters}
	\renewcommand{\arraystretch}{1.3}
	\begin{tabular}{c|c}
		\firsthline
		\textbf{Parameters}              & \textbf{Values}  \\
		\hline   
		ST density ($\lambda_s$) & $2\times10^{-4} $      \\
		Secondary link distance ($d_s$)     & $40$ m\\
		Primary link distance ($d_p$)     & $300$ m\\
		Cell size ($R$)      & $500$ m      \\
		Pathloss exponent ($\alpha$)     & $4$     \\  
		PT power ($P_1$) & $100$ mW      \\
		Maximum ST power ($P_{2,\max}$) & $0.02$ mW \\
		Noise power ($\sigma^2$)      & $-113.97$ dBm     \\
		SINR target ($\theta$)      & $0$ dB     \\
		Average delay threshold ($D_{\max}$)      & $3.5$ time slots/packet     \\
		\lasthline
	\end{tabular}	
	\label{system_params}
\end{table}

In this section we evaluate the secondary throughput as a function of the two variables $(q_2, P_2)$ within their feasible region that satisfies the delay constraints of the primary user. The primary delay and the feasible region boundary are also presented, showing the impact of the priority-based protocol design on the network performance. The values of the parameters are given in Table~\ref{system_params}. 

\begin{figure}[h]
	\includegraphics[scale=0.45]{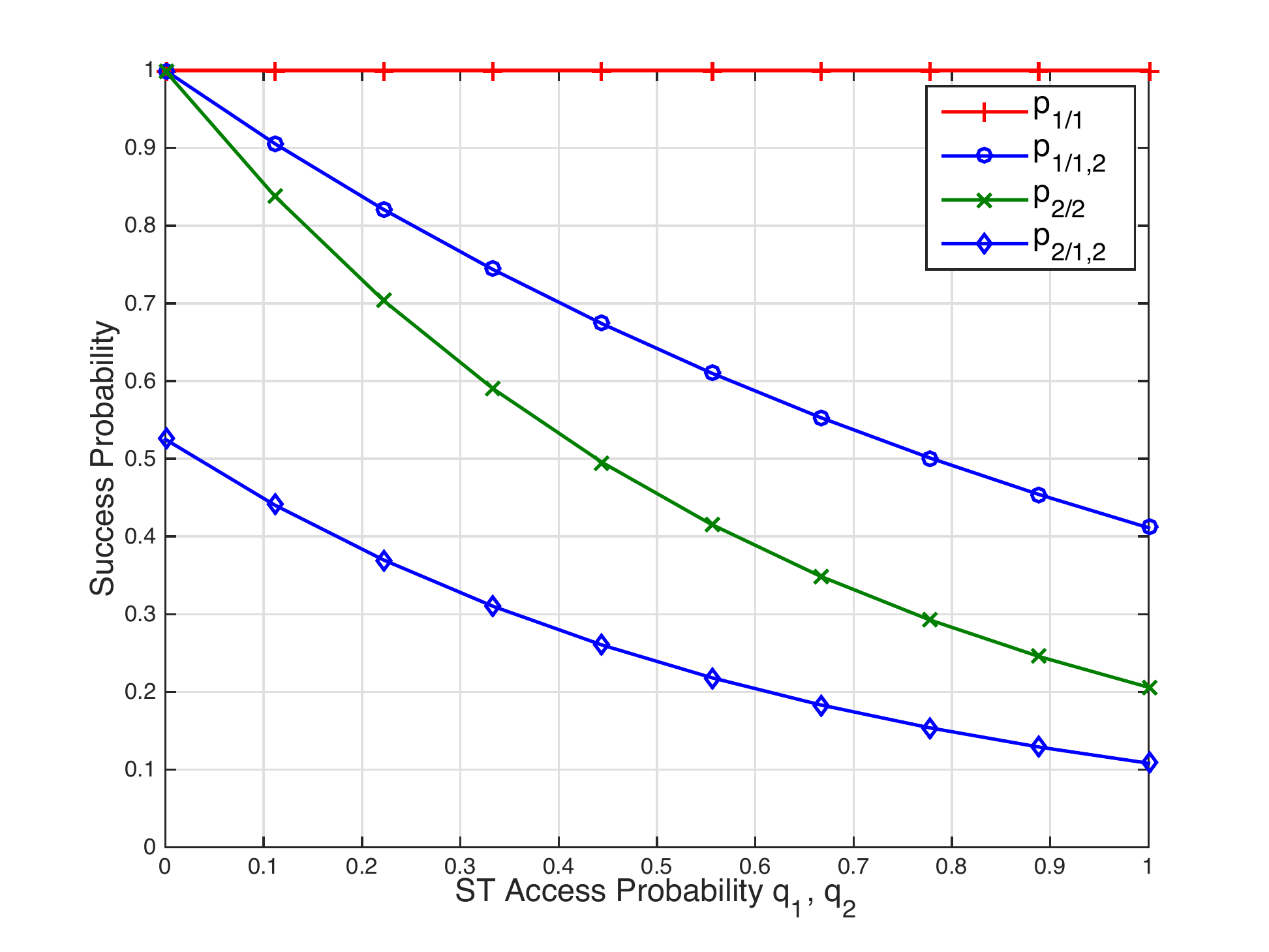}
	\centering
	\caption{Success probabilities of the primary and secondary transmissions vs. ST access probability. $p_{1/1}$ is constant, $p_{2/2}$ is a function of $q_1$, $p_{1/1,2}$ and $p_{2/1,2}$ are functions of $q_2$. The ST power is set to $P_2=0.01$ mW. }
	\label{fig:success}
\end{figure}

In Fig.~\ref{fig:success}, we plot the success probabilities $p_{1/1}$, $p_{1/1,2}$, $p_{2/2}$ and $p_{2/1,2}$ as a function of the ST access probability $q_1$ or $q_2$ for ST power set to $P_2=0.01$ mW. The numerical values are obtained from \eqref{eq:p11}, \eqref{eq:p112}, \eqref{eq:p22} and \eqref{eq:p212}, respectively. Recall that $p_{1/1}$ is a constant value, $p_{1/1,2}$ and $p_{2/1,2}$ depend only on $q_2$,  and $p_{2/2}$ depends only on $q_1$. As expected, when the secondary network is active, the success probabilities decrease rapidly with $q_1$ and $q_2$ increasing, as a result of the increased interference level. 

\subsection{General Case}

\begin{figure}[!ht]
	\centering
	\subfigure[$\lambda=0.3$, $M=1$]{
		\includegraphics[scale=0.45]{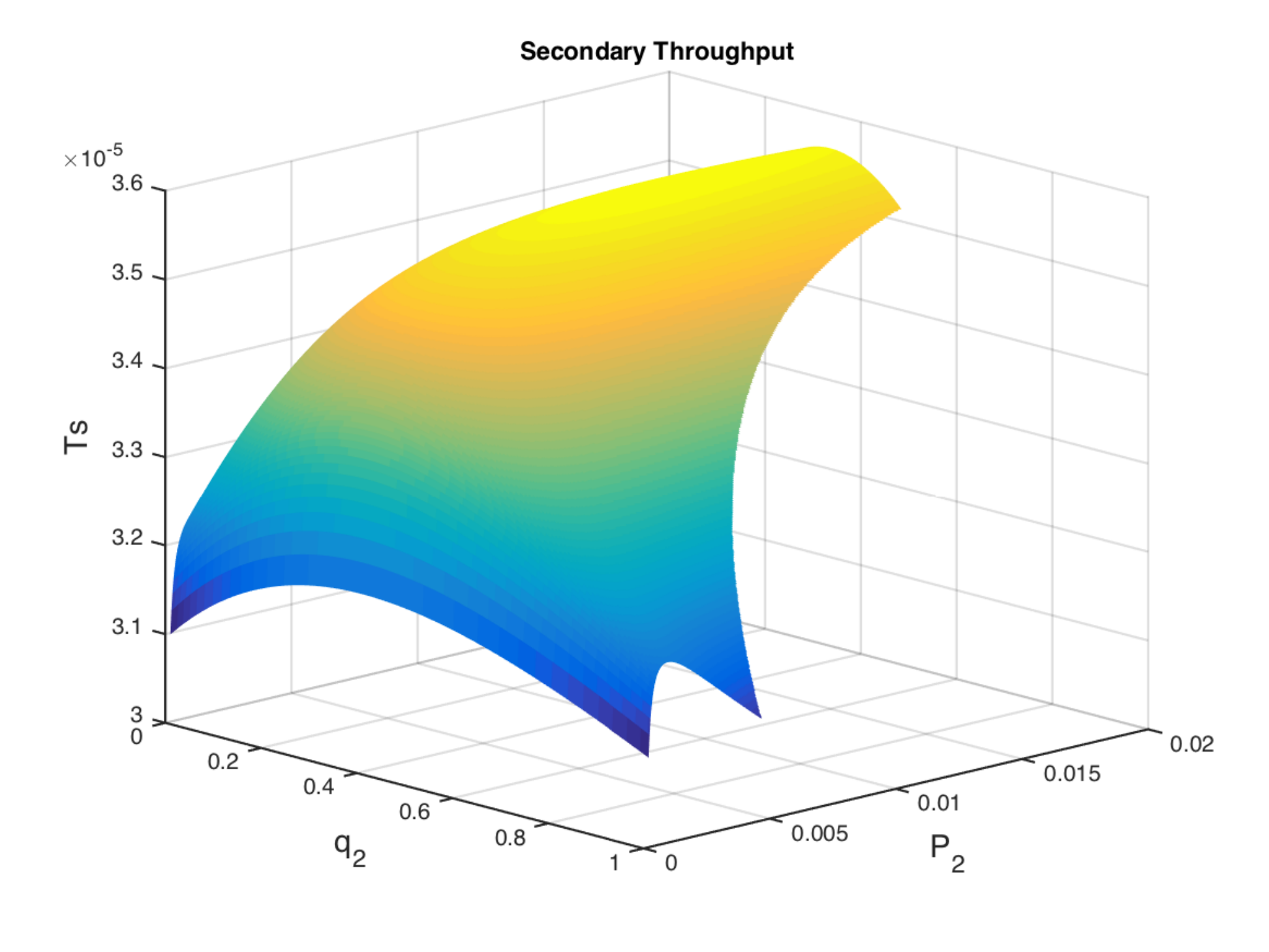}
		\label{fig:SU_throughput_0p3_M1}
	}
	\subfigure[$\lambda=0.3$, $M=3$]{
		\includegraphics[scale=0.45]{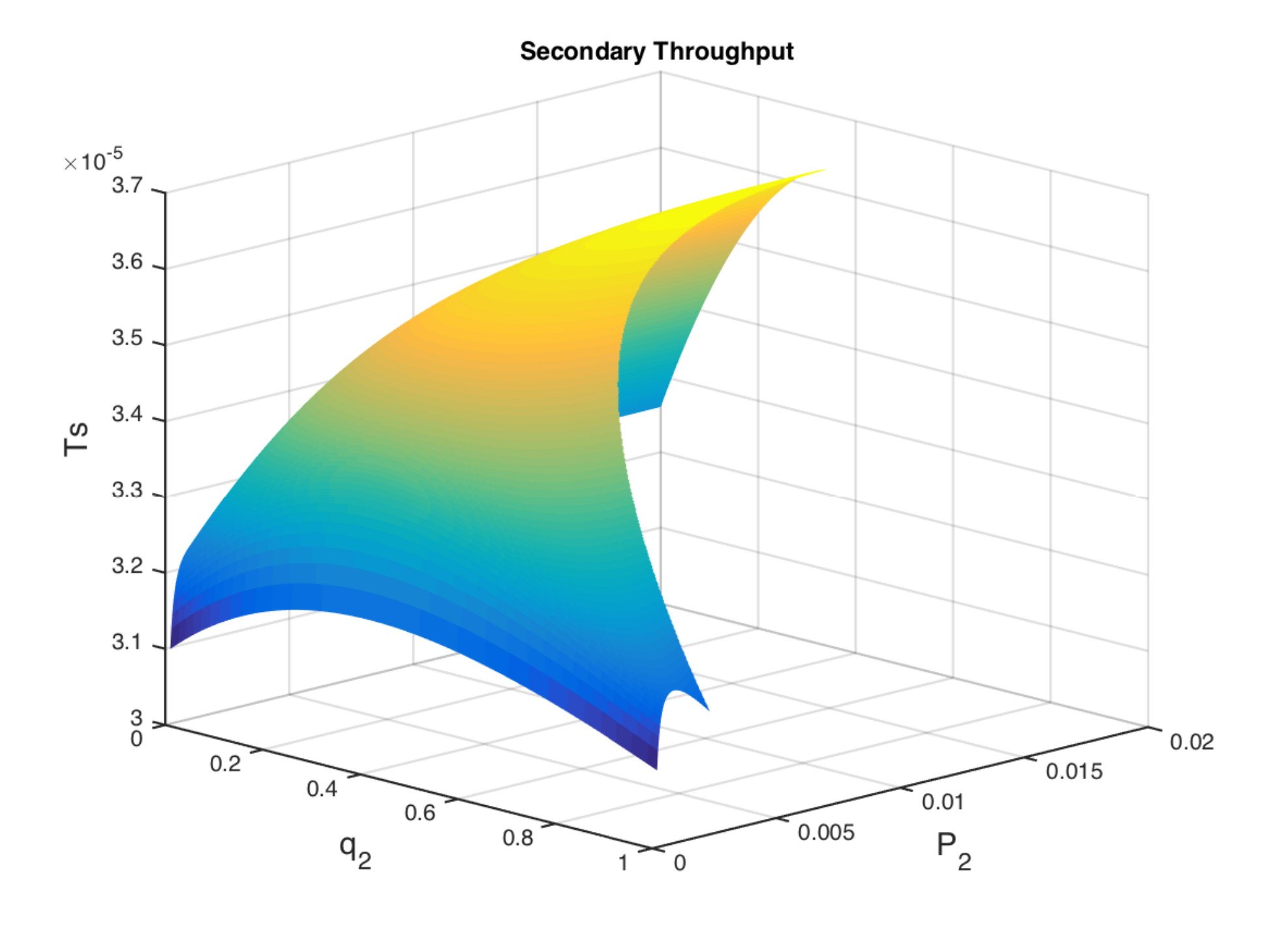}
		\label{fig:SU_throughput_0p3_M3}
	}
	\caption{Secondary throughput vs. $(q_2, P_2)$ under primary delay constraints. $\lambda=0.3$.}
	\label{fig:SU_throughput_1}
\end{figure}

\begin{figure}[!ht]
	\centering
	\subfigure[$\lambda=0.7$, $M=1$]{
		\includegraphics[scale=0.45]{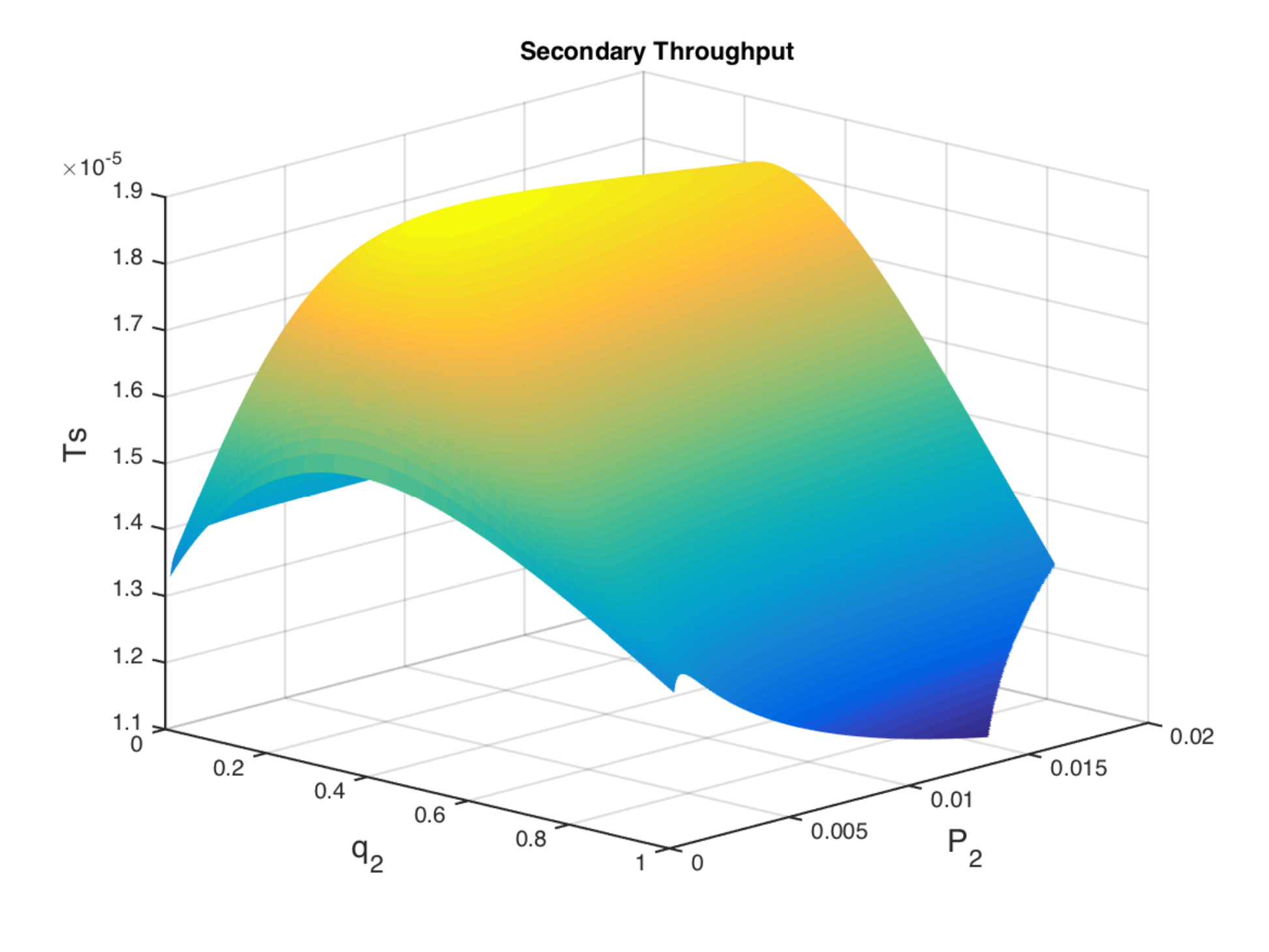}
		\label{fig:SU_throughput_0p6_M1}
	}
	\subfigure[$\lambda=0.7$, $M=3$]{
		\includegraphics[scale=0.45]{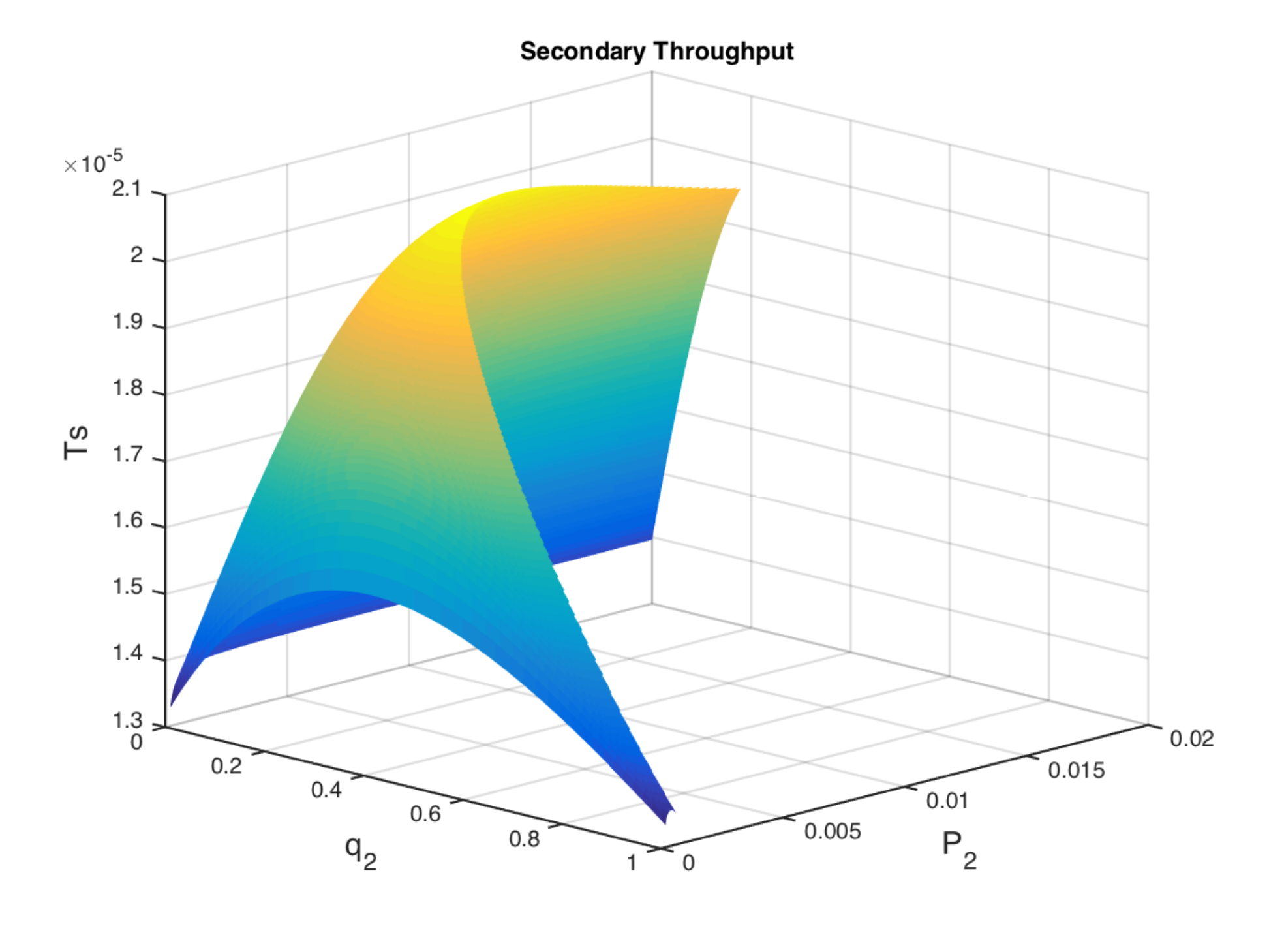}
		\label{fig:SU_throughput_0p6_M3}
	}
    \caption{Secondary throughput vs. $(q_2, P_2)$ under primary delay constraints. $\lambda=0.7$.}
	\label{fig:SU_throughput_2}
\end{figure}

\begin{figure}[h]
	\includegraphics[scale=0.45]{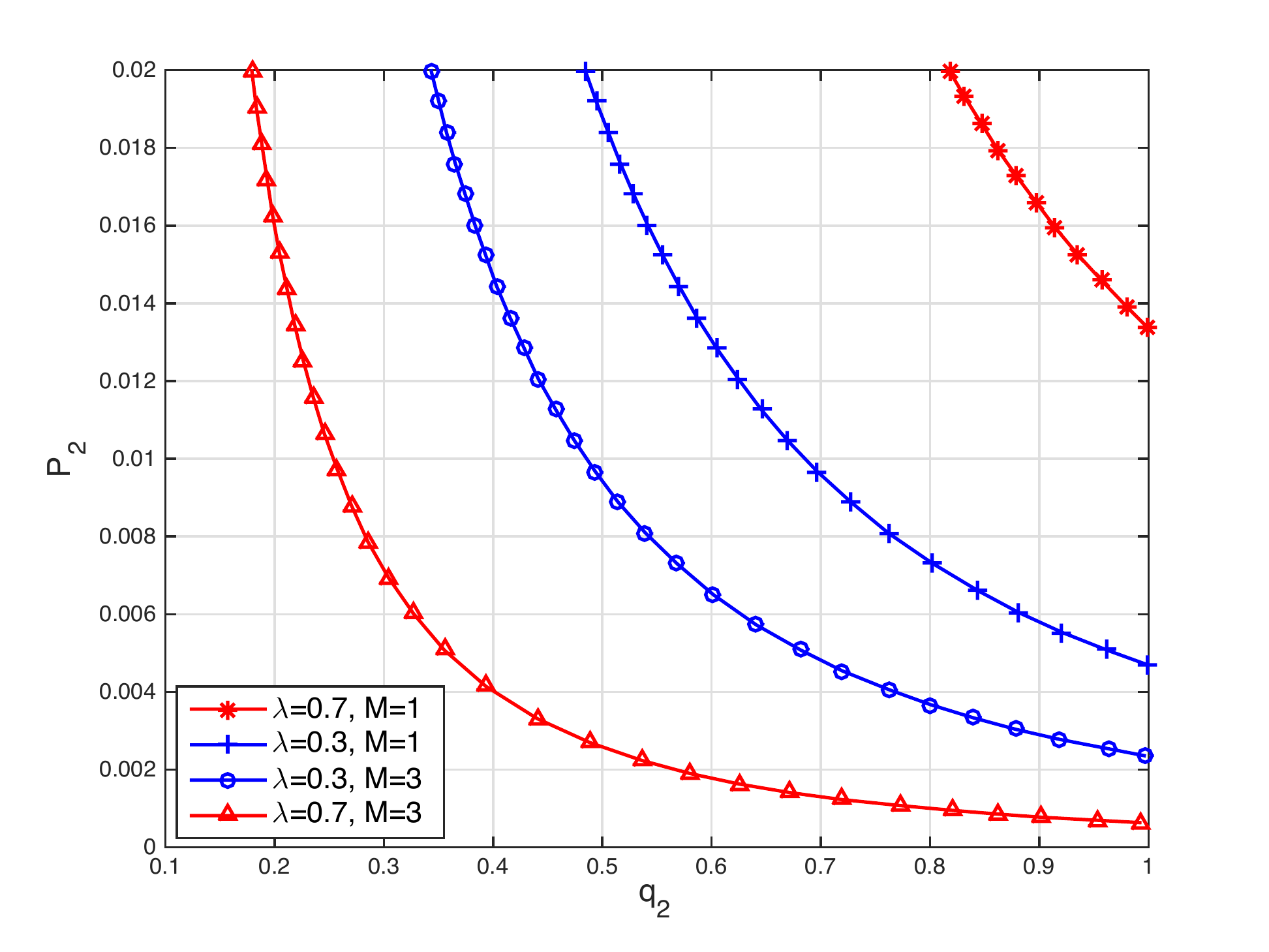}
	\centering
	\caption{The boundary of the feasible region of $(q_2, P_2)$ with $M=\{1, 3\}$ and $\lambda=\{0.3, 0.7\}$. Below each curve is the feasible region $\mathcal{R}_{\mathcal{F}}$ with respective values of $\lambda$ and $M$. }
	\label{fig:feasible_region}
\end{figure}

\begin{table}[t]
	\centering
	\caption{Optimal SU Setting with Throughput Maximization}
	\renewcommand{\arraystretch}{1.3}
	\begin{tabular}{ | c | c | c |c | c|}
		\hline
		$\lambda$ & $M$ & $q_2^{*}$ & $P_2^{*}(\text{mW})$ & $T_s^{*} (\times10^{-5})$ \\ \hline
		\multirow{2}{*}{$0.7$} & $1$ & $0.29$ & $0.0062$ & $1.87$ \\  \cline{2-5}
		& $3$ & $0.304$ & $0.0081$ & $2.08$   \\ \hline
		\multirow{2}{*}{$0.5$} & $1$ & $0.323$ & $0.0094$ & $2.76$ \\ \cline{2-5}
		& $3$ & $0.344$ & $0.012$ & $2.91$   \\ \hline
		\multirow{2}{*}{$0.3$} & $1$ & $0.349$ & $0.0124$ & $3.57$ \\ \cline{2-5}
		& $3$ & $0.377$ & $0.0177$ & $3.63$   \\ 
		\hline
	\end{tabular}
	\label{table_data}
\end{table}

In Fig.~\ref{fig:SU_throughput_1} and Fig.~\ref{fig:SU_throughput_2}, we plot the secondary throughput under the primary delay constraints. The values of $T_s$ are obtained from \eqref{eq:sd_thpt_proba} within the feasible region of $(q_2, P_2)$ defined in \eqref{eq:def_RF_general}. The results are presented with congestion threshold $M=\{1,3\}$ and the packet arrival rate $\lambda=\{0.3, 0.7\}$. Knowing that $\mu_2=p_{1/1}=0.9997$, we choose $\lambda<0.9997$ in order to satisfy the queue stability condition. 

Our first remark is that the secondary throughput is not a monotonic function of $q_2$ and $P_2$. There exists an optimal point that gives the maximum $T_s$ among the feasible choices of $(q_2, P_2)$.  We also observe a ceiling effect, i.e. once $P_2$ reaches a certain level, e.g., $P_2=0.006$ in Fig.~\ref{fig:SU_throughput_1}, $T_s$ has very small variation with respect to variations of $P_2$. This result implies that in order to have throughput gains, the necessary power for the secondary transmission should actually be quite low. Thus, the condition we used in Lemma~\ref{theo_optimal_qo} is validated.

Comparing the subfigures we observe that larger $M$ provides higher potential improvement for the secondary throughput, as the secondary links are more likely to be active.  In order to validate our conclusion,
in Table.~\ref{table_data} we give the numerical values of the optimal solution $(q_2^*, P_2^*)$ as well as the maximum SU throughput achieved with different $\lambda$ and $M$. We can see that for the same $\lambda$, larger $M$ increases the maximum achievable secondary throughput, and also the optimal values of $q_2$ and $P_2$ are higher. 

Furthermore, in Fig.~\ref{fig:feasible_region} we draw the boundary of the feasible region $\mathcal{R}_{\mathcal{F}}$ for the four cases with $M=\{1, 3\}$ and $\lambda=\{0.3, 0.7\}$ respectively. The possible values of $(q_2, P_2)$ that satisfy the primary delay constraints are situated below each plot. We observe that, larger $M$ leads to more restricted feasible region, because in this case the congestion control is weaker, thus causes higher primary delay. Interestingly, we remark that the feasible region with $\lambda=0.7$ and $M=1$ is larger than that with $\lambda=0.3$ and $M=1$. This means that for the same values of $(q_2, p_2)$, the primary average delay obtained with $\lambda=0.7$ is actually smaller than the case with $\lambda=0.3$. This is mainly due to the benefits of the congestion control in protecting the primary node transmission when the queue size is large. With high packet arrival rate, i.e., $\lambda=0.7$, the probability of having $Q>M$ is very high, thus the STs will remain silent with high probability. In that case both the queueing delay and the transmission delay of the primary user will be reduced. 

\begin{figure}[!ht]
	\hspace*{-0.5cm}
	\includegraphics[scale=0.51]{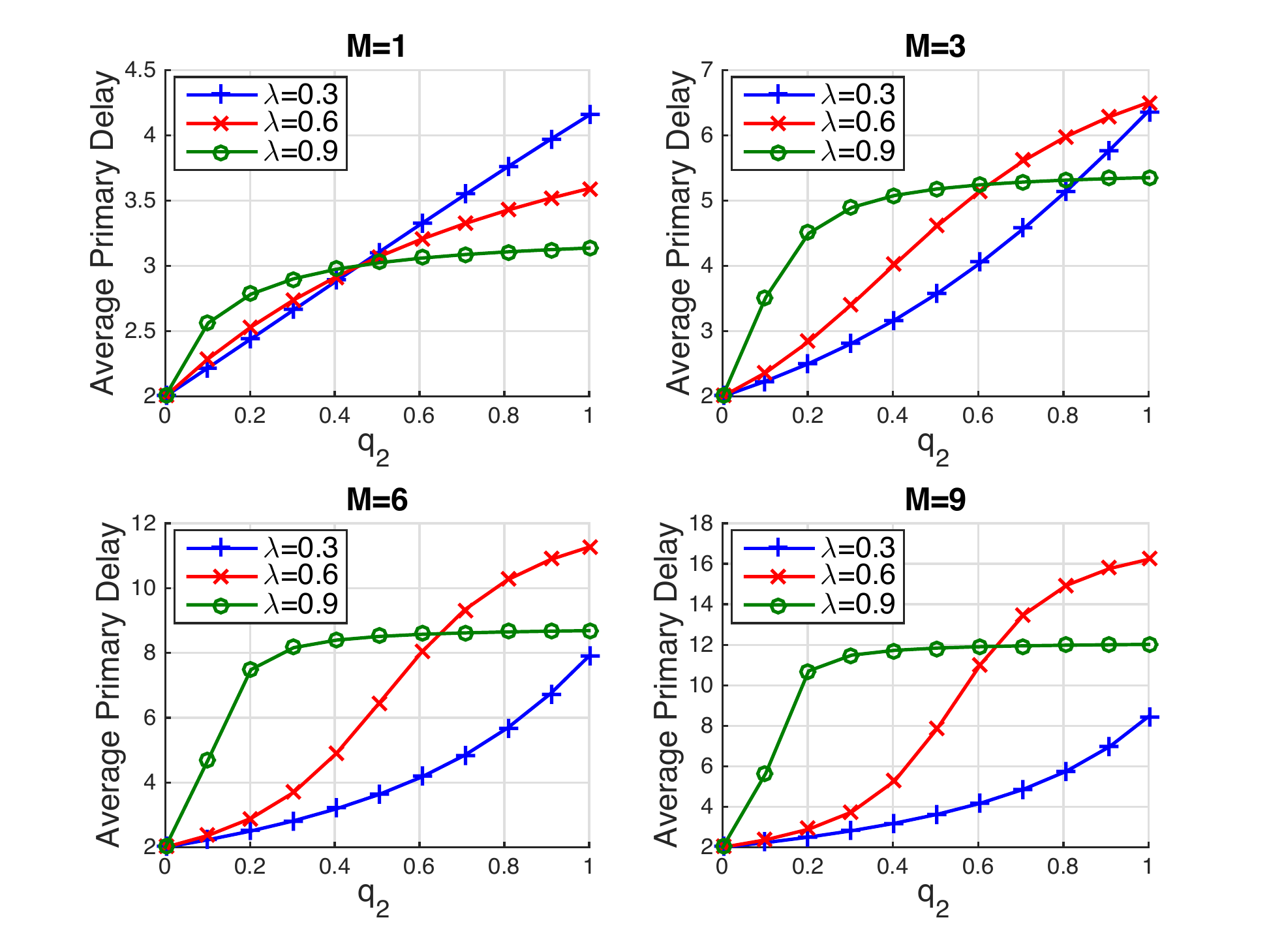}
	\centering
	\caption{Primary average delay $\bar{D}_p$ vs. ST access probability $q_2$. $P_2=0.01$ mW. $M=\{1,3,6,9\}$. $\lambda=\{0.3, 0.6, 0.9\}$}
	\label{fig:delay_vs_q2}
\end{figure}

In order to further understand the influence of $M$ and $\lambda$ on the primary delay, in Fig.~\ref{fig:delay_vs_q2} we plot the primary average delay as a function of the ST access probability $q_2$ for different values of $M$ and $\lambda$. The ST power is set to $P_2=0.01$ mW. Note that all the results are obtained with $\lambda<p_{1/1}$ in order to satisfy the queue stability condition. We have the following observations:
\begin{enumerate}
	\item With $q_2$ increasing, which corresponds to the case of the PT service rate $\mu_1$ decreasing, the primary delay increases rapidly at first, then saturating. The higher the arrival rate $\lambda$ is, the lower saturated delay it gives.
	\item When $q_2$ is relatively small, which means relatively high service rate $\mu_1$, the primary delay is higher in the case with higher arrival rate $\lambda$. However, when $q_2$ is relatively high, depending on the value of $M$, this trend can be contrasting, e.g., in the case with $M=1$, when $q_2>0.46$, the primary average delay is lower than in the case with higher $\lambda$.
	\item Larger $M$ results in higher primary average delay, thus requires higher service rate $\mu_1$ (smaller $q_2$) in order to satisfy the delay constraints. 
\end{enumerate}

The main takeaway messages we have from these results are:
\begin{enumerate}
	\item With larger $M$, the maximum secondary throughput is higher. However, larger $M$ put tighter constraints on the feasible values of $(q_2, P_2)$. 
	\item With higher arrival rate $\lambda$ at the PT, both the ST access probability and transmit power should be set to be lower. By doing so, the primary user achieves higher service rate, thus the queue size decreases faster, which in turn gives higher chance to the STs to transmit during the next time slot.
	\item  When the primary user is very sensitive to the delay, smaller $M$ is more beneficial in order to increase the primary transmission rate.
\end{enumerate}

\subsection{Case with no Congestion Control ($M=\infty$)}
\begin{figure}[ht!]
	\centering
	\subfigure[ $\lambda=0.3$]{
		\includegraphics[scale=0.45]{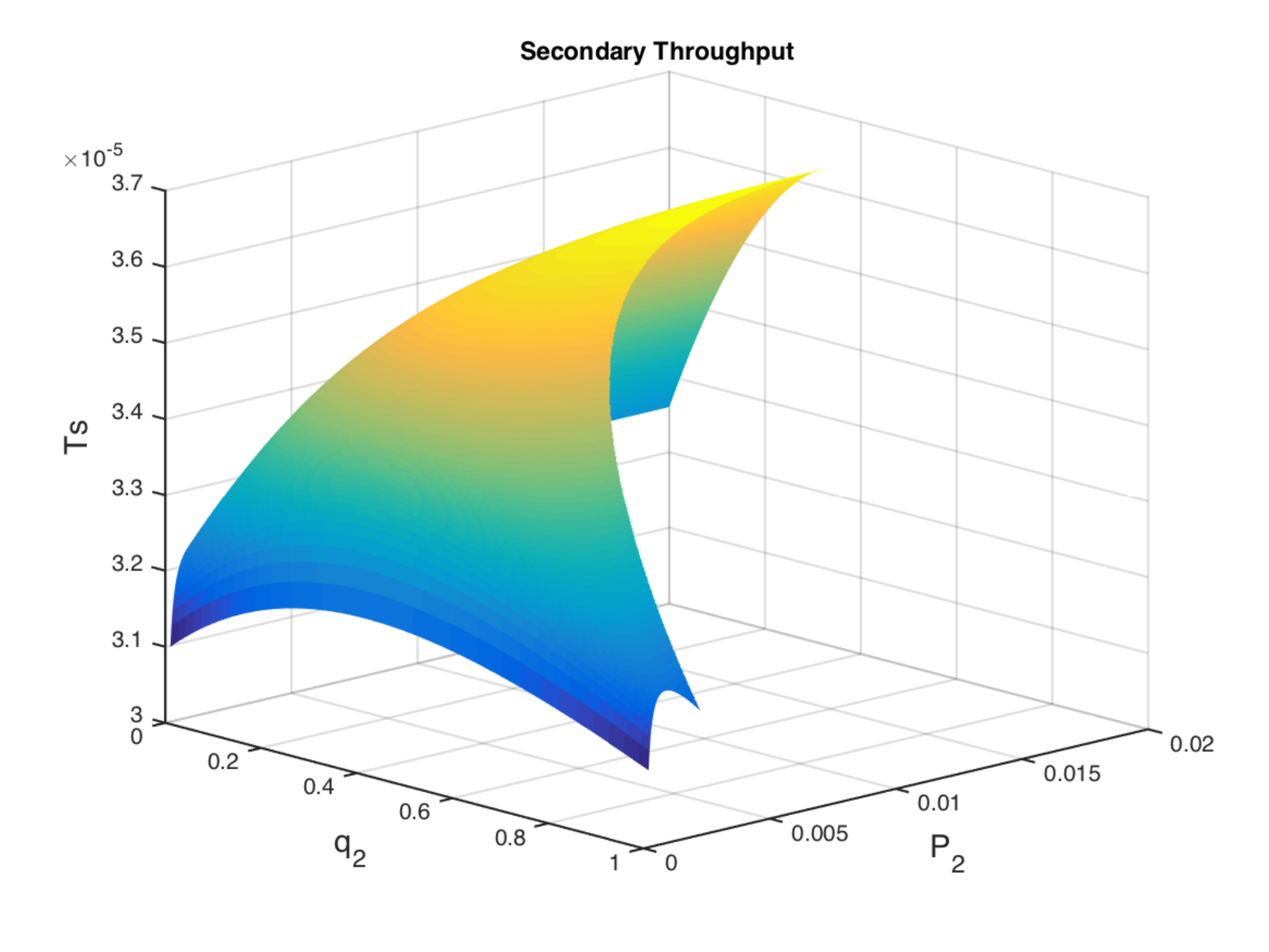}
		\label{fig:thpt_simp1}
	}
	\subfigure[ $\lambda=0.7$]{
	\includegraphics[scale=0.45]{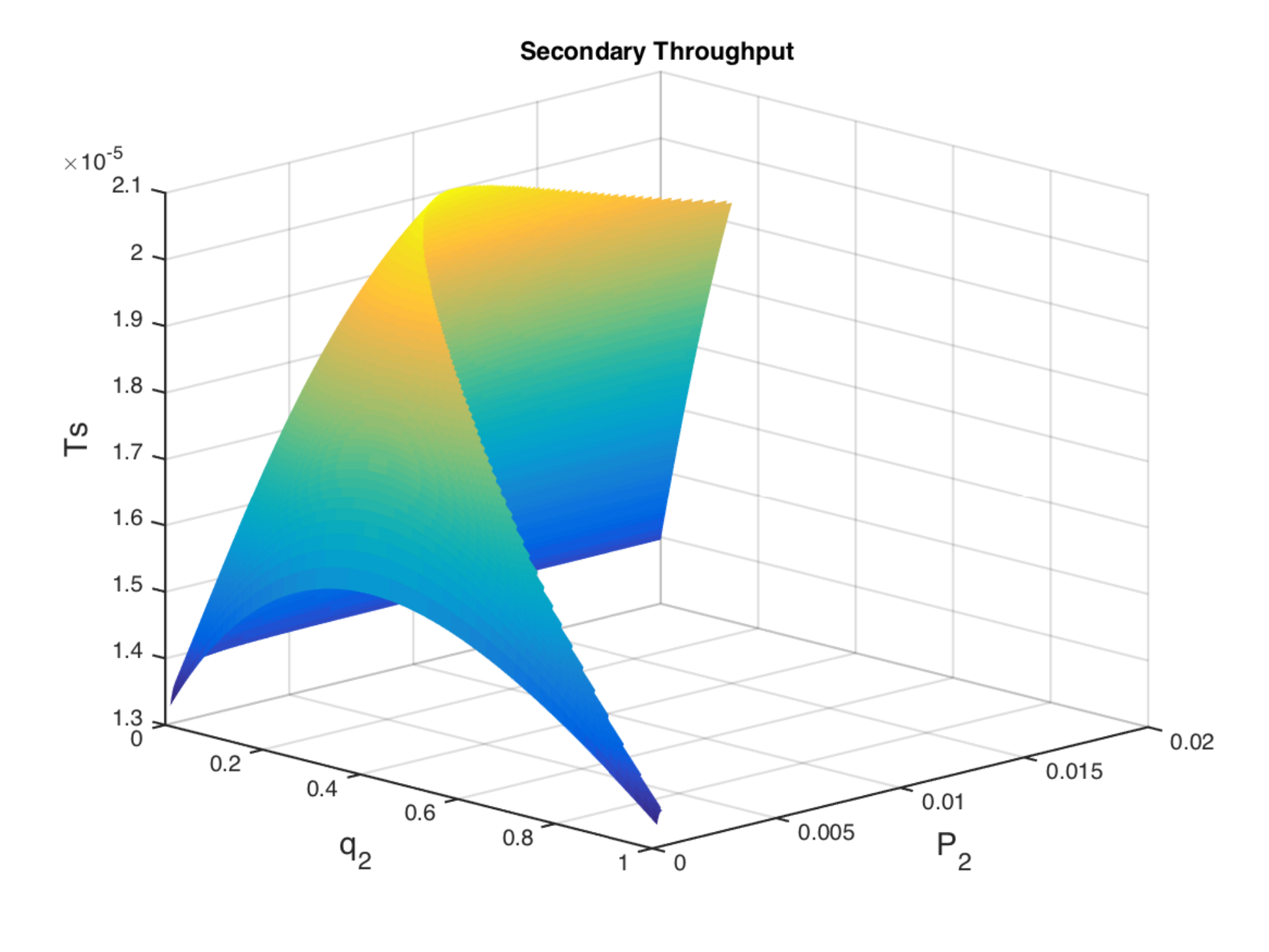}
	\label{fig:thpt_simp2}
	}
	\caption{Secondary throughput vs. $(q_2, P_2)$ under primary delay constraints, in the simplified case without congestion control.}
	\label{fig:thpt_simp}
\end{figure}

\begin{figure}[h]
	\includegraphics[scale=0.45]{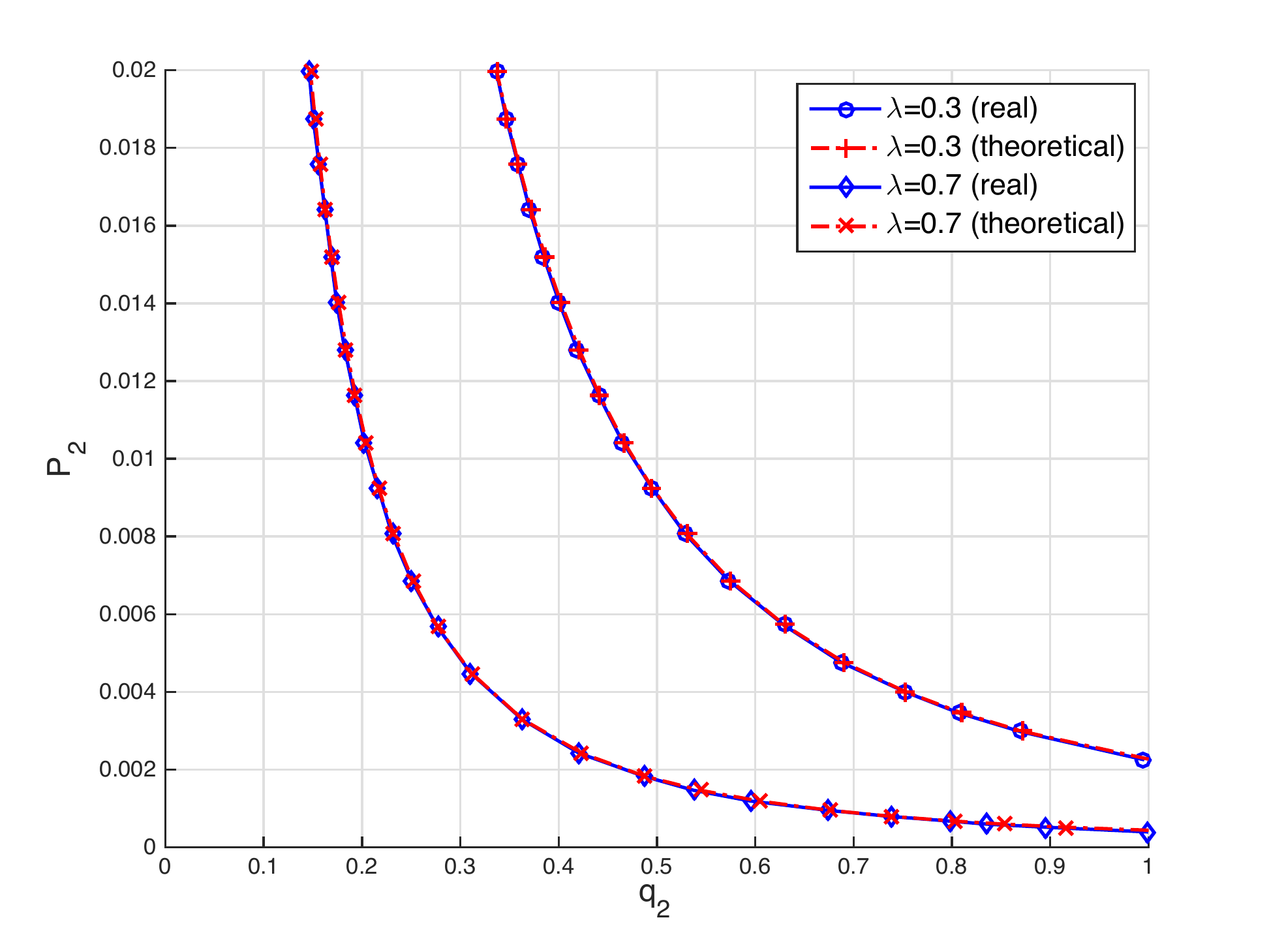}
	\centering
	\caption{The boundary of the feasible region of $(q_2, P_2)$ with $\lambda=\{0.3, 0.7\}$. The real values are obtained with exhaustive search.}
	\label{fig:feasible_region_theo}
\end{figure}

\begin{figure}[h]
	\includegraphics[scale=0.45]{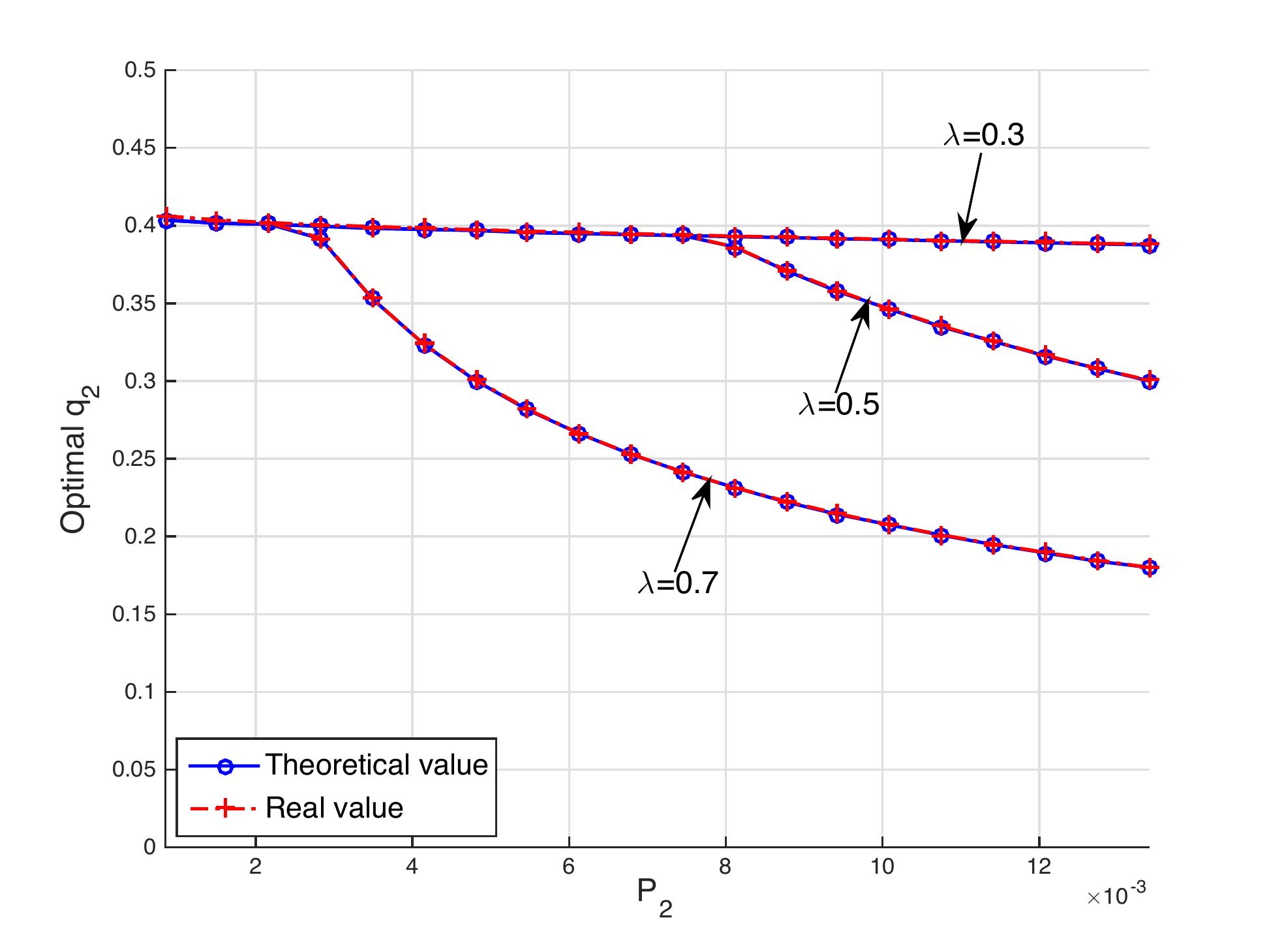}
	\centering
	\caption{Optimal access probability $q_2^*$ vs. $P_2$. The real values are obtained with exhaustive search.}
	\label{fig:optimal_q2}
\end{figure}

In Fig.~\ref{fig:thpt_simp}, we plot the secondary throughput under the primary delay constraints in the case without congestion control. The results are presented for $\lambda=\{0.3, 0.7\}$.
The evolution of the secondary throughput follows the same trend as observed in the general case presented in Fig.~\ref{fig:SU_throughput_1} and Fig.~\ref{fig:SU_throughput_2}.

Fig.~\ref{fig:feasible_region_theo} shows the theoretical boundary of the feasible region $\mathcal{R}_{\mathcal{F}}$ derived in Lemma~\ref{theo_feasible_region} in comparison to the real boundary obtained by exhaustive search of the feasible values of $(q_2, P_2)$ under primary delay constraints. The results confirm the accuracy of our theoretical analysis on the feasible region of $(q_2, P_2)$.

Fig.~\ref{fig:optimal_q2} shows the optimal access probability $q_2^*$ obtained with Theorem~\ref{theo_optimal_q2} in comparison to the the real optimal values obtained by exhaustive search of $q_2$ that maximizes the secondary throughput with respect to the primary delay constraints. This illustrate the accuracy of our analytical results in Theorem~\ref{theo_optimal_q2}. Another observation is that with $\lambda=0.3$, $q_2^*$ has values close to $0.4$. When $\lambda$ is higher, $q_2^*$ declines rapidly with $P_2$ after $P_2$ reaches a certain value. This result is expected because above a certain value of $P_2$, the optimal $q_2$ is equal to the maximum feasible value of $q_2$ which is at the boundary of $\mathcal{R}_{\mathcal{F}}$.

\section{Conclusions} \label{sec:conclusions}
This paper investigated a delay-constrained shared access network following a stochastic geometry approach. We proposed a priority-based protocol with congestion control and studied the throughput of the secondary network and the primary average delay, as well as the impact of the protocol design parameters on the throughput and delay performance. For the case without congestion control, we derived in closed-form the optimal access probability of the secondary node in terms of maximizing the throughput of the secondary network under primary delay constraints. The main contribution of this work was to analyze the performance of a shared access network with priorities using tools from stochastic geometry, as well as to extend prior work on throughput optimization in shared access networks to the case with primary delay constraints.

\appendices
\section{Proof of Proposition~\ref{propi_p212} }
\label{appen1}
According to the definition of the success probability, for the typical active secondary pair $i$, we have 
\begin{align}
p_{2/1, 2}&=\mathbb{P}\left[\rm SINR_i >\theta\mid \mathcal{T}=\{x_0 \cup \Phi_a^2\}\right]    \nonumber\\
& =\mathbb{P}\left[\frac{P_2|h_{i,i}|^2 d_s^{-\alpha}}{\sigma^2\!+\!\!\!\! \sum\limits_{j\in \Phi_a^2\backslash\left\{i\right\}} \!\!\! P_{2}|h_{j,i}|^2 d_{j,i}^{-\alpha}\!+\!P_{1}|h_{0,i}|^2 d_{0,i}^{-\alpha}} >\theta\right]  \nonumber\\
&\mathop{=}\limits^{(a)} \exp\!\!\left[\!-\theta d_s^\alpha\left(\!\frac{\sigma^2}{P_2} +\frac{P_{1}}{P_2}|h_{0,i}|^2 d_{0,i}^{-\alpha} \!+\!\!\!\!\!\!\!\!\sum\limits_{j\in \Phi_a^2\backslash\left\{i\right\}} \!\!\!\!\!\!\! |h_{j,i}|^2 d_{j,i}^{-\alpha}\!\!\right)\!\right]       \nonumber\\
& \mathop{=}\limits^{(b)} \exp\left(-\frac{\theta \sigma^2 d_s^\alpha}{P_2}\right)   \mathbb{E}\left[\frac{1}{1+\frac{P_{1}}{P_2} \theta d_s^\alpha d_{0,i}^{-\alpha}}\right] \mathcal{L}_{I_s}(\theta d_s^\alpha) .
\label{eq:p212_derivation}
\end{align}
Here, $(a)$ follows from $|h_{i,i}|^2\sim \exp(1)$. $(b)$ follows from  $|h_{0,i}|^2\sim \exp(1)$, and the expectation is over $d_{0,i}$. $\mathcal{L}_{I_s}(s)=\mathbb{E}\left[\exp\left(-s \sum\limits_{j\in \Phi_a^2\backslash\left\{i\right\}} \!\!\!\!\! |h_{j,i}|^2 d_{j,i}^{-\alpha}\right)\right]$ is the Laplace transform of interference coming from active STs with normalized transmit power.

With the help of the approximation $\mathbb{E}\left[\frac{1}{1+\frac{\kappa}{d_{0,i}^{\alpha}}}\right]\simeq \frac{1}{1+\frac{\kappa^{2/\alpha}}{\mathbb{E}[d_{0,i}]^2}}$ in \cite{jeff_d2d}, the second term in \eqref{eq:p212_derivation} becomes
\begin{equation}
\mathbb{E}\left[\frac{1}{1+\frac{P_{1}}{P_2} \theta d_s^\alpha d_{0,i}^{-\alpha}}\right] \simeq\frac{1}{1+\frac{d_s^2}{\mathbb{E}[d_{0,i}]^2}\left(\theta\frac{P_1}{P_2}\right)^{\frac{2}{\alpha}}}.
\label{eq:expectation}
\end{equation}
Depending on the distance from the PT to the active SRs, different SRs experience different interference levels caused by the primary transmission. 
The expectation of $d_{0,i}$ is over all the possible locations of the typical SR inside the network region $\mathcal{C}$. 

The distribution of the active SRs depends on the locations of their associated STs, which follows a homogeneous PPP with intensity $q_2 \lambda_s$. For an arbitrary active SR, it can be approximately seen as uniformly distributed on the disk $\mathcal{C}$ with radius $R$. Hence, the pdf of the distance from the typical SR to the origin of $\mathcal{C}$, denoted by $d_1$, is given by
\begin{equation}
f_{d_1}(r)= \left\{
\begin{array}{rcl}
 &\!\!\!\!\!\!\!\frac{2 r}{R^2}  & \text{if}~~ 0\leq r \leq R\\
&\!\!\!\!\!\!\!0   & \text{else}.
\end{array} \right.
\end{equation}
\begin{figure}[h]
	\includegraphics[scale=0.2]{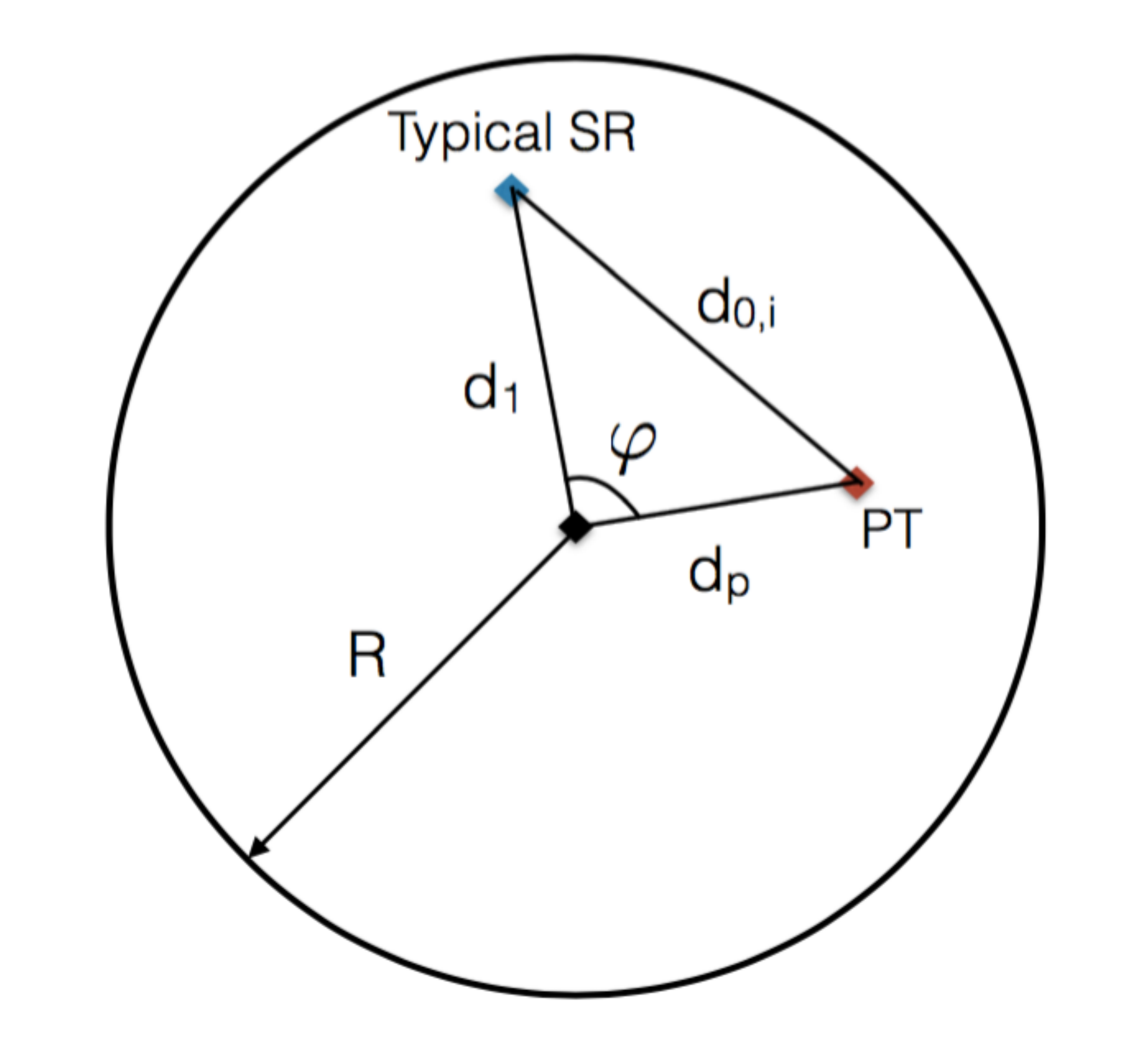}
	\centering
	\caption{Geographical locations of the PT and the typical SR on the network region $\mathcal
		{C}$ with radius $R$.}
	\label{fig:cosine}
\end{figure}
The distance from the PT to the origin is $d_p$. As shown in Fig.~\ref{fig:cosine}, 
using the law of cosine, we have the distance between the PT and the typical SR given by
\begin{equation}
d_{0,i}=\sqrt{d_1^2+d_p^2-2d_1 d_p \cos \varphi},
\end{equation}
where $\varphi$ is a random variable uniformly distributed in $[0, 2\pi]$. Averaging over $d_1$ and $\varphi$, we have the expectation of the distance $d_{0,i}$ given by 
\begin{equation}
\mathbb{E}[d_{0,i}]=\int_{0}^{2\pi}\frac{1}{2\pi}\int_{0}^{R}\frac{2 r}{R^2}\sqrt{r^2+d_p^2-2r d_p \cos \varphi}\text{d}r\text{d}\varphi. 
\label{eq:approxi_d0i}
\end{equation}
The third term in \eqref{eq:p212_derivation} is the Laplace transform of interference coming from nodes in $\Phi_a^2\backslash\left\{i\right\}$ with intensity $q_2 \lambda_s$. From existing results on the interference distribution in Poisson networks \cite{haenggi_interference}, we have 
\begin{equation}
\mathcal{L}_{I_s}(\theta d_s^\alpha)  = \exp\left[-\frac{\pi q_2\lambda_s d_s^2 \theta^{\frac{2}{\alpha}}}{\sinc(2/\alpha)}\right] .
\label{eq:ppp_interference}
\end{equation}
Substituting \eqref{eq:expectation} and \eqref{eq:ppp_interference} in \eqref{eq:p212_derivation}, together with \eqref{eq:approxi_d0i}, we have
\begin{equation*}
p_{2/1,2}\simeq\exp\left[-\frac{\pi q_2\lambda_s d_s^2 \theta^{\frac{2}{\alpha}}}{\sinc(2/\alpha)}\right] \frac{\exp\left(-\frac{\theta \sigma^2 d_s^\alpha}{P_2}\right)}{1+\frac{d_s^2}{\mathbb{E}[d_{0,i}]^2}\left(\theta\frac{P_1}{P_2}\right)^{\frac{2}{\alpha}}},
\end{equation*}
where $\mathbb{E}[d_{0,i}]$ is given in \eqref{eq:approxi_d0i}. Proposition~\ref{propi_p212} is obtained.

\section{Proof of Lemma~\ref{lemma_dtmc} }
\label{appen2}
From the DTMC described in Fig.~\ref{fig:dtmc}, we obtain the following balance equations.
\begin{equation*}
\lambda \pi(0) = (1-\lambda) \mu_1 \pi(1) \Leftrightarrow \pi(1) = \frac{\lambda}{(1-\lambda) \mu_1} \pi(0)
\end{equation*}
\begin{align*}
\left[\lambda (1-\mu_1) +(1-\lambda)\mu_1\right]\pi(1) = \lambda \pi(0)+(1-\lambda)\mu_1 \pi(2) \\
\Leftrightarrow \pi(2) = \frac{\lambda^2 (1-\mu_1)}{(1-\lambda)^2 \mu_1^2} \pi(0).
\end{align*}
Summarizing, for $1 \leq i \leq M$ we have that
\begin{equation*}
\pi(i) = \frac{\lambda^i (1-\mu_1)^{i-1}}{(1-\lambda)^i \mu_1^i} \pi(0),
\end{equation*}
and for $i>M$ we obtain
\begin{equation*}
\pi(i) = \frac{\lambda^{i} (1-\mu_1)^M (1-\mu_2)^{i-M-1}}{(1-\lambda)^{i} \mu_1^{M} \mu_2^{i-M}} \pi(0).
\end{equation*}
Knowing that
\begin{equation} \label{eq:sum}
\sum_{i=0}^{\infty} \pi(i)=1,
\end{equation}
combined with the previous expressions, when $\lambda\neq\mu_1$, the probability that the queue is empty is given by 
\begin{equation}
\pi(0)= \frac{(\mu_1 - \lambda)(\mu_2 - \lambda)}{\mu_1 \mu_2 - \lambda \mu_1 - \lambda \left[\al\right]^M (\mu_2 - \mu_1) }. 
\label{eq:p0_first}
\end{equation}
A special case is when $\lambda=\mu_1$. Denote $g(\lambda)$ and $h(\lambda)$ the nominator and the denominator of $\pi(0)$. Since $g(\mu_1)=h(\mu_1)=0$, \eqref{eq:p0_first} is no longer valid. By using l'H\^{o}pital's rule, we have 
\begin{equation}
\pi(0)= \lim\limits_{\lambda\rightarrow\mu_1}\frac{g'(\lambda)}{h'(\lambda)}=\frac{\mu_2-\mu_1}{\mu_1+(\mu_2-\mu_1)\frac{M+1-\mu_1}{1-\mu_1}}.
\label{eq:p0_second}
\end{equation}
Combining the two cases with $\lambda\neq \mu_1$ and $\lambda=\mu_1$, we have \eqref{eq:Pempty} in Lemma~\ref{lemma_dtmc}.

The condition that the DTMC is aperiodic irreducible Markov chain, which implies that the queue is stable, is $\lambda<\mu_2$. Since $\pi(0)$ is a positive probability, we have an additional condition $0<\pi(0)<1$ that $\lambda$ must satisfy. We consider the following cases:
\begin{itemize}
	\item If $\lambda<\mu_1 \Rightarrow\al <1$,  the denominator $h(\lambda)>\mu_2(\mu_1-\lambda)$. Then we have $\pi(0)<\frac{\mu_2-\lambda}{\mu_2}=1-\frac{\lambda}{\mu_2}<1$. It is also obvious that $\pi(0)>0$. 
	\item If $\lambda=\mu_1$, from \eqref{eq:p0_second} we have $0<\pi(0)<1$. 
	\item If $\mu_1<\lambda<\mu_2 \Rightarrow\al >1$, we have $g(\lambda)<0$. As for the denominator $h(\lambda)$, it can be proven that
	\begin{align*}
	\left(\frac{1-\lambda}{1-\mu_1}\right)^M \frac{\mu_2-\lambda}{\mu_2-\mu_1} &<1 <\left(\frac{\lambda}{\mu_1}\right)^{M+1} \\
	\Longrightarrow\mu_1(\mu_2-\lambda) &< \lambda \left[\al\right]^M (\mu_2-\mu_1)\\
	\Longrightarrow h(\lambda)&<0. 
	\end{align*}
	Thus we have $\pi(0)=\frac{g(\lambda)}{h(\lambda)}>0$.
	From $\al >1$ we also know that $h(\lambda)<\mu_2(\mu_1-\lambda)<0$, then we have $\pi(0)<1-\frac{\lambda}{\mu_2}<1$. 
\end{itemize}
Since in the three cases $0<\pi(0)<1$ is always verified, we obtain the necessary and sufficient condition that the queue is stable when $\lambda<\mu_2$.

\section{Proof of Lemma~\ref{lemma_p1m} }
\label{appen3}
From the results in Lemma~\ref{lemma_dtmc}, when $\lambda<\mu_2$ and $\xi\triangleq \al \neq 1$, we have
\begin{align}
\mathbb{P}[1\leq Q\leq M]& =\sum\limits_{i=1}^{M} \pi(i)  =\frac{\pi(0)}{1-\mu_1} \sum\limits_{i=1}^{M} \left[\frac{\lambda (1-\mu_1)}{(1-\lambda)\mu_1}\right]^{i}   \nonumber \\
&=\frac{\pi(0)}{1-\mu_1}  \frac{\frac{\lambda (1-\mu_1)}{(1-\lambda)\mu_1}-\left[\frac{\lambda (1-\mu_1)}{(1-\lambda)\mu_1}\right]^{M+1}}{1-\frac{\lambda (1-\mu_1)}{(1-\lambda)\mu_1}}  \nonumber \\
&= \frac{\pi(0)\lambda \left(1-\xi^{M}\right)}{\mu_1-\lambda}  \nonumber \\
&=\frac{\lambda \left(1-\xi^{M}\right)(\mu_2 - \lambda)}{\mu_1 \mu_2 - \lambda \mu_1 - \lambda \xi^M (\mu_2 - \mu_1) }.
\end{align}
We also have 
\begin{align}
\mathbb{P}[Q >M]& =\sum\limits_{i=M}^{\infty} \pi(i) =1-\sum\limits_{i=0}^{M} \pi(i)  \nonumber \\
&=\frac{\lambda \xi^M  (\mu_1-\lambda)}{\mu_1 \mu_2 - \lambda \mu_1 - \lambda \xi^M (\mu_2 - \mu_1)}.
\end{align}

\section{Proof of Theorem~\ref{lemma_delay} }
\label{appen4}
From the results in Lemma~\ref{lemma_dtmc}, we have the average size of the queue at the PT given by
\begin{align}
\bar{Q}&=\sum\limits_{i=1}^{\infty}i \pi(i) = \sum\limits_{i=1}^{M}i \pi(i) +\sum\limits_{i=1}^{\infty}(M+i) \pi(M+i)    \nonumber \\
&= \sum\limits_{i=1}^{M}i \pi(i) +M\sum\limits_{i=1}^{\infty} \pi(M+i) + \sum\limits_{i=1}^{\infty}i \pi(M+i) .
\label{eq:qavg}
\end{align}

When $\lambda<\mu_2$ and $F \triangleq \al \neq 1$, the first term can be derived as follows.
\begin{align}
 \sum\limits_{i=1}^{M}i \pi(i) &= \sum\limits_{i=1}^{M} i \pi(0) \frac{\lambda^i (1-\mu_1)^{i-1}}{(1-\lambda)^i \mu_1^i}    \nonumber \\
 &=\frac{\pi(0)\lambda }{(1-\lambda)\mu_1} \sum\limits_{i=1}^{M} i\left[\frac{\lambda (1-\mu_1)}{(1-\lambda)\mu_1}\right]^{i-1}    \nonumber \\
 &=\frac{\pi(0)\lambda }{(1-\lambda)\mu_1} \sum\limits_{i=1}^{M} \left(\left[\frac{\lambda (1-\mu_1)}{(1-\lambda)\mu_1}\right]^{i} \right)^{\prime}   \nonumber \\
 &=\frac{\pi(0)\lambda }{(1-\lambda)\mu_1} \frac{M \xi^{M+1}-\xi^M(M+1) +1 }{\left(1-\frac{\lambda (1-\mu_1)}{(1-\lambda)\mu_1}\right)^2}   \nonumber \\
&=\frac{\lambda(1-\lambda)\mu_1 \frac{\mu_2-\lambda}{\mu_1-\lambda}\left[M \xi^{M+1}-\xi^M(M+1) +1\right]}{\mu_1\mu_2 -\lambda\mu_1 -\lambda \xi^M  (\mu_2-\mu_1)}.
\label{eq:term1}
\end{align}   
For the second term, with the help of \eqref{eq:Pr_M}, we have
\begin{align}
M\sum\limits_{i=1}^{\infty} \pi(M+i) &=M \left(1-\mathbb{P}[Q>M]\right)   \nonumber \\
&=\frac{\lambda(\mu_1-\lambda)M \xi^M}{\mu_1\mu_2 -\lambda\mu_1 -\lambda \xi^M  (\mu_2-\mu_1)}.
\label{eq:term2}
\end{align}
For the third term, with the help of \eqref{eq:pi_bis}, we have
\begin{align}
\sum\limits_{i=1}^{\infty}i \pi(M+i)& = \frac{\xi^M \pi(0)\lambda}{(1-\lambda)\mu_2}\sum\limits_{i=1}^{\infty} i\left[\frac{\lambda (1-\mu_1)}{(1-\lambda)\mu_1}\right]^{i-1}    \nonumber \\
&= \frac{\xi^M \pi(0)\lambda}{(1-\lambda)\mu_2}  \frac{1}{\left[1-\frac{\lambda (1-\mu_2)}{(1-\lambda)\mu_2}\right]^2}      \nonumber\\
&= \frac{\lambda (1-\lambda)\mu_2\frac{\mu_1 - \lambda}{\mu_2-\lambda}\xi^M }{\mu_1\mu_2 -\lambda\mu_1 -\lambda \xi^M  (\mu_2-\mu_1)}.
\label{eq:term3}
\end{align}
Substituting \eqref{eq:term1}, \eqref{eq:term2} and \eqref{eq:term3} in \eqref{eq:qavg}, we have
\begin{equation}
\bar{Q}=\frac{N_1+N_2}{\mu_1\mu_2 -\lambda\mu_1 -\lambda \xi^M  (\mu_2-\mu_1) },
\label{eq:qavg_final}
\end{equation}
where 
\begin{equation}
N_1=\lambda(1-\lambda)\mu_1 \frac{\mu_2-\lambda}{\mu_1-\lambda}\left[M \xi^{M+1}-\xi^M(M+1) +1\right],
\end{equation}
and 
\begin{equation}
N_2=\xi^M \lambda(\mu_1-\lambda)\left[M+\frac{(1-\lambda)\mu_2}{\mu_2-\lambda}\right].
\end{equation}

From the definition of the primary average delay in Section~\ref{sec:def_delay} and the expression of the average queue size $\bar{Q}$ given in \eqref{eq:qavg_final},  we have
\begin{align}
\bar{D}_p
&=\frac{\bar{Q}}{\lambda}+\frac{\mathbb{P}[ Q\neq0]}{\mathbb{P}[1\leq Q\leq M]\mu_1+\mathbb{P}[ Q> M]\mu_2 }    \nonumber \\
&=\frac{\bar{Q}}{\lambda} +\frac{\mu_2-\lambda-\xi^M(\mu_2-\mu_1)}{\left(1-\xi^M \right)(\mu_2 - \lambda) \mu_1+\xi^M  (\mu_1-\lambda) \mu_2}.   
\end{align}
 Theorem~\ref{lemma_delay} is obtained.

\section{Proof of Lemma~\ref{theo_optimal_qo} }
\label{appen5}
Define $\kappa_1=\frac{\pi d_s^2 \theta^{2/\alpha}}{\sinc(2/\alpha)}$, $\kappa_2=\frac{\pi d_p^2 \left(\theta\frac{P_2}{P_1}\right)^{2/\alpha}}{\sinc(2/\alpha)}$, and $\kappa_{12}=\frac{1}{1+\frac{d_s^2}{\mathbb{E}[d_{0,i}]^2}\left(\theta\frac{P_1}{P_2}\right)^{\frac{2}{\alpha}}}$ constant parameters related to the network setting, \eqref{eq:thp_SU_noM_2} becomes
\begin{align}
T_{s}& = c^{*}+\frac{\lambda}{p_{1/1}} \frac{q_2 \exp(-q_2 \lambda_s \kappa_1)\cdot \kappa_{12}-c^{*}}{\exp(-q_2 \lambda_s \kappa_2)}  \nonumber\\
&=  c^{*}\!+\! \frac{\lambda \kappa_{12}}{p_{1/1}} \left\{q_2\exp\left[q_2\lambda_s(\kappa_2-\kappa_1)\right]\!-\!\frac{c^{*}}{\kappa_{12}}\exp(q_2\lambda_s\kappa_2)\right\}. \nonumber
\end{align}
Define $c_{12}^{*}=\frac{c^{*}}{\kappa_{12}}=q_1^{*} p_{2/2}(q_1^{*})\left[1+\frac{d_s^2}{\mathbb{E}[d_{0,i}]^2}\left(\theta\frac{P_1}{P_2}\right)^{\frac{2}{\alpha}}\right]$, we need to find the optimal value of $q_2$ that maximizes $T_s$ with respect to $q_2\in[0,1]$, i.e., 
\begin{equation}
q_2^{\text{o}}=\argmax_{q_2\in[ 0,1]}~ q_2\exp\left[q_2\lambda_s(\kappa_2-\kappa_1)\right]- c_{12}^{*}\exp(q_2\lambda_s\kappa_2).
\label{eq:global_optim}
\end{equation}

We define the following objective function $f(x)=x\exp\left[x\lambda_s(\kappa_2-\kappa_1)\right]- c_{12}^{*}\exp(x\lambda_s\kappa_2)$ with $x\in[0,1]$. 
First, $f(x)$ is not for sure a concave function. Secondly, maximizing $f(x)$ depends on whether $(\kappa_1-\kappa_2)$ is positive or negative. Taking the first order derivative of $f(x)$, we have 
\begin{equation}
f' (x)=e^{x\lambda_s\kappa_2}\left\{e^{-x\lambda_s\kappa_1} \left[1-x\lambda_s (\kappa_1-\kappa_2)\right]-c_{12}^{*}\lambda_s\kappa_2\right\}.
\label{eq:derivative}
\end{equation}

When $\frac{P_2}{P_1}<\left(d_s/d_p\right)^{\alpha}$, $\kappa_1-\kappa_2>0$ holds. Obviously $f' (x)$ decreases with $x$, and we have $\lim\limits_{x\rightarrow-\infty}f' (x)=+\infty$ and $\lim\limits_{x\rightarrow+\infty}f' (x)=-\infty$. For $x\in(-\infty, +\infty)$, the only critical point of $f(x)$ is the global optimal (maximum) point, which can be easily obtained by using the first order optimality condition.
Considering that $x$ is bounded by $x\in[0,1]$, we can find the optimal point in the following cases.  
\begin{itemize}
	\item If $f'(1)<f'(0)<0$, $f(x)$ monotonically decreases in $x\in[0,1]$. The optimal point is at $x^{o}=0$.
	\item If $f'(0)>f'(1)>0$, $f(x)$ monotonically increases in $x\in[0,1]$. The optimal point is at $x^{o}=1$.
	\item If $f'(0)>0>f'(1)$, the optimal point is at $x^{o}$ such that $f'(x)=0$. 
\end{itemize}  
From \eqref{eq:derivative}, the first order optimality condition gives 
\begin{align}
&e^{-x\lambda_s\kappa_1} \left[1-x\lambda_s (\kappa_1-\kappa_2)\right]-c_{12}^{*}\lambda_s\kappa_2=0.  \nonumber \\
\Longrightarrow ~~&e^{x\lambda_s\kappa_1}=-\frac{\kappa_1-\kappa_2}{c_{12}^{*} \kappa_2}x +\frac{1}{c_{12}^{*} \lambda_s \kappa_2}.
\label{eq:exponential_poninomial}
\end{align}
For a general type of equation $p^{ax+b}=cx+d$, where $x$ is the variable to be solved and $a$, $b$, $c$, $d$, $p$ are constant, when $p>0$ and $a,c\neq 0$, the solution by using the Lambert $W$ function is
\begin{equation}
x=-\frac{W\left(-\frac{a\ln p}{c} p^{b-\frac{ad}{c}}\right)}{a\ln p}-\frac{d}{c}.
\label{lambert}
\end{equation}
Solving \eqref{eq:exponential_poninomial} with the help of the Lambert W function, combined with the condition $q_2\in[0,1]$, we have the solution to \eqref{eq:global_optim} when $\frac{P_2}{P_1}<\left(d_s/d_p\right)^{\alpha}$, given by
\begin{equation}
q_2^{\text{o}}=\min\left\{\!\left[-\frac{W\left(\frac{\lambda_s \kappa_{1}\kappa_{2} c_{12}^{*}}{\kappa_{1}-\kappa_{2}} e^{\frac{\kappa_1}{\kappa_{1}-\kappa_{2}}}\right)}{\lambda_s \kappa_1}+\frac{1}{\lambda_s(\kappa_{1}-\kappa_{2})}\right]^{+}\!\!\!, 1\right\}\!, \nonumber
\end{equation}
where $[z]^{+}=\max\{z, 0\}$.

When $\frac{P_2}{P_1}\geq \left(d_s/d_p\right)^{\alpha}$, $f'(x)$ is not a monotonic function of $x$. Therefore, $f(x)$ may have more than one critical points, depending on the shape of $f(x)$ with different network parameters. Here, we disregard the case where $\frac{P_2}{P_1}\geq \left(d_s/d_p\right)^{\alpha}$ in order to have tractable analysis on the optimization problem.

Combining these results, we have the Lemma~\ref{theo_optimal_qo}.

\section{Proof of Lemma~\ref{theo_feasible_region} }
\label{appen6}
The feasibility region $\mathcal{R}_\mathcal{F}$ is defined by the intersection of the queue stability condition and the queue size constraint. $(q_2, P_2)$ should satisfy
\begin{enumerate}
	\item $\lambda< \mu_1$;
	\item $\frac{1-\lambda}{\mu_1-\lambda} +\frac{1}{\mu_1}<D_{\max}$.
\end{enumerate}
From the first condition, we have
\begin{equation}
\lambda< p_{1/1,2}(q_2) \Rightarrow q_2< \frac{\ln (P_{1/1}/\lambda)}{ \lambda_s \kappa_2}.
\label{eq:condition_1}
\end{equation}
From the second condition, we have
\begin{equation}
D_{\max} \mu_1^2 -[(D_{\max} -1)\lambda+2] \mu_1 +\lambda<0 
\end{equation}
The solution to the inequality is 
\begin{equation}
q_2<\frac{\ln (P_{1/1}/\eta_1)}{\lambda_s \kappa_2} ~~\text{or} ~~ q_2>\frac{\ln (P_{1/1}/\eta_2)}{\lambda_s \kappa_2} 
\label{eq:condition_2}
\end{equation}
where $\eta_1=\frac{(D_{\max}-1)\lambda+2+\sqrt{(D_{\max}-1)^2\lambda^2-4\lambda +4}}{2D_{\max}}$ and $\eta_2=\frac{(D_{\max}-1)\lambda+2-\sqrt{(D_{\max}-1)^2\lambda^2-4\lambda +4}}{2D_{\max}}$.

Knowing that $\eta_2<\lambda$ always holds, the intersection of \eqref{eq:condition_1} and \eqref{eq:condition_2} gives
\begin{equation}
q_2< \min\left\{ \frac{\ln (P_{1/1}/\lambda)}{ \lambda_s \kappa_2},  \frac{\ln (P_{1/1}/\eta_1)}{ \lambda_s \kappa_2}\right\}.
\end{equation}
The feasible region of $(q_2, P_2)$ is thus defined by 
\begin{equation}
\mathcal{R}_{\mathcal{F}}=\left\{(q_2, P_2): q_2< \min\left\{ \frac{\ln (P_{1/1}/\lambda)}{ \lambda_s \kappa_2},  \frac{\ln (P_{1/1}/\eta_1)}{ \lambda_s \kappa_2}\right\}\right\}.
\end{equation}

\section{Proof of Theorem~\ref{theo_optimal_q2} }
\label{appen7}
When $\frac{P_2}{P_1}< \left(d_s/d_p\right)^{\alpha}$, from \eqref{eq:derivative} we have $f'(0) >0$. Knowing that $f'(x)$ decreases with $x$,  $f(x)$ is either an monotonically increasing function or firstly increases then decreases in $x\in[0,1]$. 

If $q_2^{o}$ obtained in \eqref{eq:optimal_q2_general} falls within the feasible region $\mathcal{R}_{\mathcal{F}}$ given in \eqref{eq:feasible_region_theo}, i.e, $q_2^{o}<\min\left\{ \frac{\ln (P_{1/1}/\lambda)}{ \lambda_s \kappa_2},  \frac{\ln (P_{1/1}/\eta_1)}{ \lambda_s \kappa_2}\right\}$ , the optimal value of $q_2$ with respect to the delay constraints is  $q_2^{o}$. Otherwise $f(x)$ is an increasing function in $\mathcal{R}_{\mathcal{F}}$, and the optimal value is the one at the feasible region boundary $\min\left\{ \frac{\ln (P_{1/1}/\lambda)}{ \lambda_s \kappa_2},  \frac{\ln (P_{1/1}/\eta_1)}{ \lambda_s \kappa_2}\right\}$.

Combining the two cases, we obtain Theorem~\ref{theo_optimal_q2}.

\bibliographystyle{IEEEtran}
\bibliography{thesis2}

\end{document}